\documentclass[reprint,amsmath,amssymb,aip,showpacs,pra]{revtex4-1}
\usepackage{graphicx}
\usepackage{dcolumn}
\usepackage{bm}
\usepackage[usenames,dvipsnames]{color}
\usepackage[utf8]{inputenc}
\usepackage{hyperref}
\graphicspath{{./Files/PDFs/}}

\begin{document}


\title{Compressibility effects on quasistationary vortex and transient hole patterns through vortex merger}

\author{Rupak Mukherjee}
\email{rupakmukherjee01@gmail.com; rupak@ipr.res.in}
\author{Akanksha Gupta}
\author{Rajaraman Ganesh}
\email{ganesh@ipr.res.in}
\affiliation{Institute for Plasma Research, HBNI, Bhat, Gandhinagar - 382428, India}

\begin{abstract}
The effect of compressibility in hydrodynamic vortex merging has been discussed. In the past, in incompressible limit it has been observed that the merging of a collection of intense point-like vortices arranged uniformly outside a circular vortex, can lead to quasistationary vortex patch and transient hole pattern inside the patch via nonlinear merger process. These patterns are akin to \textquoteleft vortex crystals\textquoteright. Compressibility can introduce a natural acoustic scale to the problem. We find that the natural mode is independent of the number of point-like vortices and the amplitude scales linearly with compressibility. Further it has been identified that after merging, the system exhibits oscillation at a natural frequency together with its harmonics and beats with its own harmonics. The power of the frequency is found to scale as $M^{-2}$, where $M$ is the Mach number. Also the vortex crystals formed out of the merging process are found to melt faster as compressibility is increased.
\end{abstract}

\maketitle


\section{Introduction}
Intense vortices are ubiquitous in two dimensional fluid and plasma turbulences. In case of a fully developed turbulence, several eddies and vortices of different sizes and strength form, merge and keep the energy flowing from a driving scale to viscous scales via inertial scales. The nonlinear energy transfer from one mode to another mode is mediated through the dynamics and merging of vortices, thus essentially controlling the time evolution of the fluid or plasma. Hence the vortex merger problem has been studied extensively to better understand their role in turbulence. The understanding of turbulence problem becomes even more complex when fluids can respond to another inherent natural timescale due to compressibility. Most of the fluids as well as plasmas in nature are compressible. Hence the effect of compressibility on the dynamics of fluid elements had remained a point of fundamental interest since many years. Surprisingly the effect of compressibility on vortex dynamics was ignored for quite a long time and was not stressed until the work of McCune \cite{mccune:1961}. Recently there has been a surge of literature in the area of compressible vortices and their dynamics\cite{mandella:1987, morton:1989, colonius:1991, bershader:1995, badwal:2014, shivani:2016} in which it has been shown that compressible vortices actually affect the nonlinear coupling between the modes and thus lead to several novel features which are absent in the incompressible flow.\\

In this paper we investigate the effect of compressibility on the dynamics and evolution of prearranged asymetric vortices. The model system generates interaction between vortices and holes (region of zero vorticity) with the background fluid via a complex process consisting of surface wave generation, wave breaking, vortex capture eventually leading to a long time structure formation or vortex crystals, which dissipate at later times. The results have been generalised to compressible regime in this report.


We start with a model system of uniform circular patch vortex with radius $R_{pa}$ at the centre of the simulation domain. A set of $N_{pv}$ number of intense point-like vortices are uniformly placed at a distance $R_{pv}$ outside the patch vortex. The vorticity strength of point-like vortex and that of the patch vortex are $\omega_{pv} = 10$ and $\omega_{pa} = 1$ respectively. The system is then time evolved numerically and the merging of the vortices are observed. As found earlier \cite{ganesh:2002}, we observe that such a system can sustain its pattern through the merging process and after merging the intense point-like vortices form a coherent \textquotedblleft pattern\textquotedblright~within the patch vortex. In this report we study the system with and without compressibility and try to identify the effects of compressibility on such patterns. We find that compressibility induces a natural mode to the system in which the total kinetic energy of the system starts to oscillate which in turn affects the timescale of merging. We have also identified that this natural frequency remains unaffected by a change in number of point vortices.


\subsection{Physics aspects}
A circular uniform vortex patch (also known as Rankine vortex) is known to support surface waves (also called Kelvin waves) on its boundary. When resonantly driven, these Kelvin waves can lead to nonlinear stationary states called  \textquotedblleft V-states\textquotedblright ~ \cite{deem:1978}. When driven further, the \textquotedblleft V-states\textquotedblright ~ result in \textquotedblleft filamentation\textquotedblright ~ (thin filament of vorticity emerging out of the patch vortex) instability and wavebreaking at later times. In past, the fact that, Kelvin waves are {\it{negative energy}} waves and the interaction energy between point like vortex and a {\it{nearly}} circular patch vortex is {\it{positive}} \cite{lansky:1997} had led to the prediction of quasistationary vortex and transient hole pattern \cite{ganesh:2002} in the context of pure electron plasmas which is morphologically similar to the incompressible inviscid two dimensional fluid. In this work \cite{ganesh:2002} it was observed that the sustained vortex pattern preceded with the formation of hole capture and transient hole patterns within the patch vortex. Similar observation was later reported by Swaminathan et al \cite{perlekar:2016} in the incompressible limit. The direct parallel between such an arrangement of fluid vortices and magnetised plasma column under strong axial magnetic field had been exploited in the past \cite{ganesh:2002}. All the past works were restricted to incompressible fluids and plasmas. Compressibility effects are known to introduce an acoustic time scale. To the best of our knowledge, the role of compressibility on the way the merger dynamics would evolve has not yet been clearly brought out. In this report we have identified five distinct features due to the introduction of the compressibility effects which are absent in the previous literatures which dealt with the incompressible fluids only:
\begin{itemize}
\item It generates sound waves with a fundamental frequency, its harmonics and beats with its own harmonics.
\item All the frequencies scale linearly with Mach number ($M = \frac{U_0}{C_s}$, where, $U_0$ is the maximum velocity and $C_s$ is the sound speed in the fluid).
\item The power of all the frequencies is found to scale as $M^{-2}$.
\item The fundamental frequency can be estimated from a simple calculation of one dimensional compressible fluid column.
\item Acoustic timescales are found to affect the merging timescale.
\item The vortex crystals melt faster due to the effect of acoustic waves generated as compared to their incompressible counterpart.
\end{itemize}


\subsection{Governing equations}
In order to address the problem mentioned above, we start with simple Navier-Stokes equations. We keep the co-efficient of viscosity ($\mu$) to be spatially independent and choose the value to be very small. The sound speed ($C_{s_0}$) is defined as $C_{s_0} = \sqrt{\frac{\bar{\gamma} P_0}{\rho_0}}$ where, $\bar{\gamma}$ is the ratio between the specific heats, $P_0$ the initial pressure and $\rho_0$ the initial density. The set of equations that we evolve with time in our code are the following.\\
\begin{eqnarray*}
&& \frac{\partial \rho}{\partial t} + \vec{\nabla} \cdot \left(\rho \vec{u}\right) = 0\\
&& \frac{\partial \vec{u}}{\partial t} + (\vec{u} \cdot \vec{\nabla}) \vec{u} = \frac{\mu}{\rho} \nabla^2 \vec{u} - C_{s_0}^2 \frac{\vec{\nabla}\rho}{\rho}
\end{eqnarray*}
where $\vec{u}$ determines the velocity of a fluid element in two dimensions. The sonic Mach number is defined as $M = \frac{U_0}{C_{s_0}}$ where $U_0$ is the maximum initial velocity of the system. It is well known that initial conditions (e.g. magnitude of density\cite{terakado:2014,bayly:1992}, pressure) affect the time evolution of compressible fluids. However, throughout this work, we set $\rho_0 = 1$ unless stated otherwise.
The effect of variation of initial magnitude of density and pressure are beyond the scope of the present work and will be addressed in a future communication.\\

The next section describes the benchmarking details of our code and the parameter details using which the whole simulation is performed. Section III provides the results obtained using the parameters in both incompressible as well as the compressible system. In section IV and V we analyse the results and try to explain the simulation results using a simple basic theoretical model. We conclude with a summary of the whole problem and the results obtained and try to outline a future direction of the problem in section VI.


\section{Simulation details}
\subsection{Benchmarking of the code}
We develop a new code Compressible Fluid Dynamics 2 Dimensional (CFD-2D), capable of handling compressible fluids, that uses pseudo-spectral technique for spatial discretisation with a standard de-aliasing by zero padding using the $3/2$ rule \cite{perlekar:2009} to simulate the above set of equations. The above equations are evaluated using velocity formulation in two dimensions with periodic boundary conditions in cartesian co-ordinates. We use the multi-dimensional FFTW libraries (v-3.3.3) \cite{FFTW3:2005} to evaluate the fourier transforms, whenever needed. The time evolution of the code is performed with Adams-Bashforth \cite{perlekar:2009}, Predictor-Corrector and Runge-Kutta 4 techniques individually and all the three techniques are found to agree in the primary benchmark results. The results in this paper are calculated using Adams-Bashforth method for temporal evolution calculation.
\subsubsection{M $\rightarrow$ 0 limit}
\begin{figure}[h!]
\begin{center}
\includegraphics[scale=0.65]{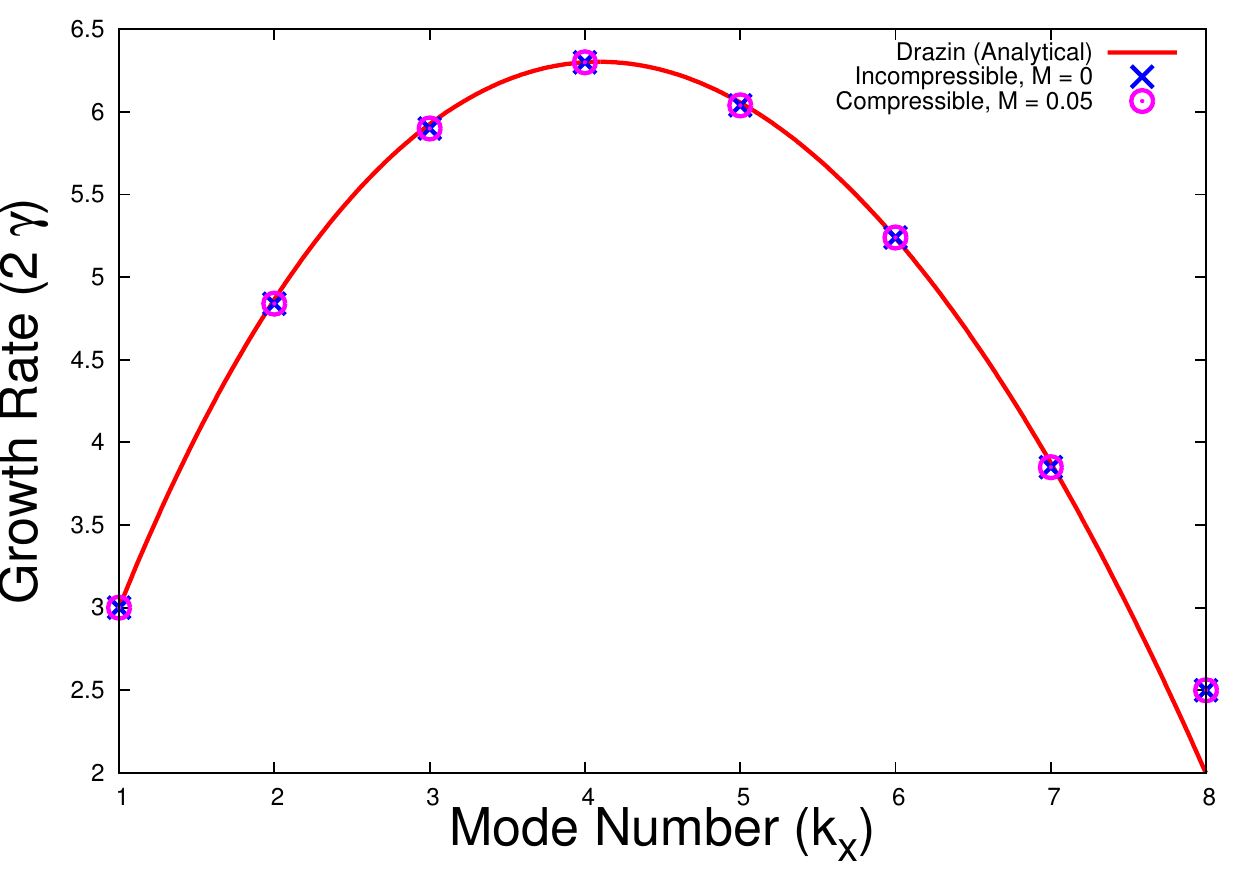}
\caption{(Color online) Growth rate ($2\gamma$) of K-H instability is plotted with mode number ($k_x$) of excitation. The red solid line is evaluated from the analytical expression obtained by Drazin. The blue line with `X' sign represents the growth rate from earlier code in vorticity formalism and the magenta line with symbol `O' represents the same from CFD-2D code with Mach number (M) = 0.05.}
\label{drazin}
\end{center}
\end{figure}

For benchmarking study we reproduce the growth rate of Kelvin-Helmholtz instability for two oppositely directed jets in an infinite domain at the incompressible limit analytically calculated earlier by Drazin \cite{drazin:1961}. We obtain the identical growth rate from CFD-2D with M = 0.05. We further confirm our results with another independently developed code that simulates the Navier-Stokes equation in vorticity formalism in the incompressible limit ($M \equiv 0$) [Fig \ref{drazin}]. We also obtain the growth rate from another existing, well-benchmarked code AG-Spect \cite{akanksha:2017} at the limit of negligible visco-elasticity. We also reproduce Kelvin-Helmholtz instability in the incompressible limit with a velocity gradient and find the same growth rate given by Ray \cite{ray:1982}.\\

\subsubsection{Finite M limit}
\begin{figure}[h!]
\begin{center}
a\includegraphics[scale=0.65]{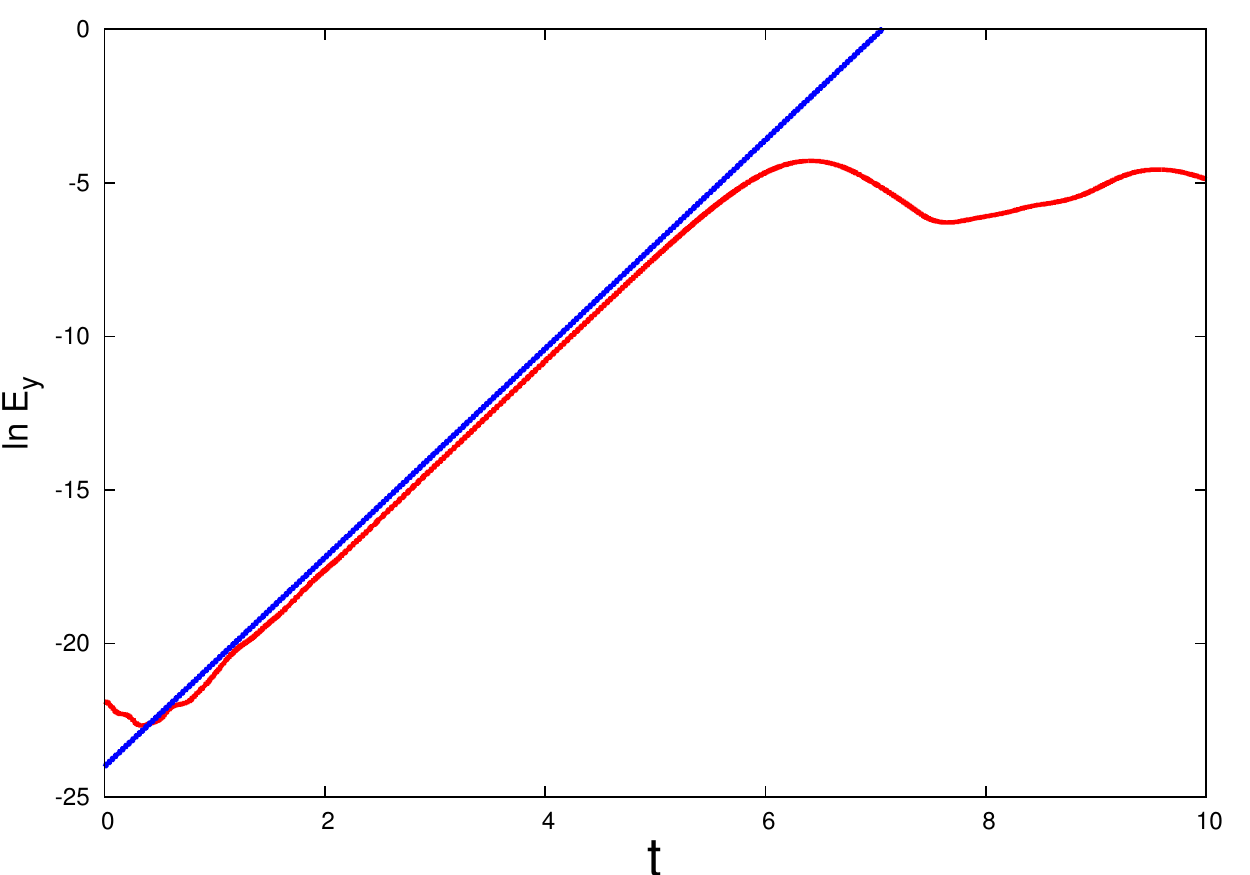}
b\includegraphics[scale=0.65]{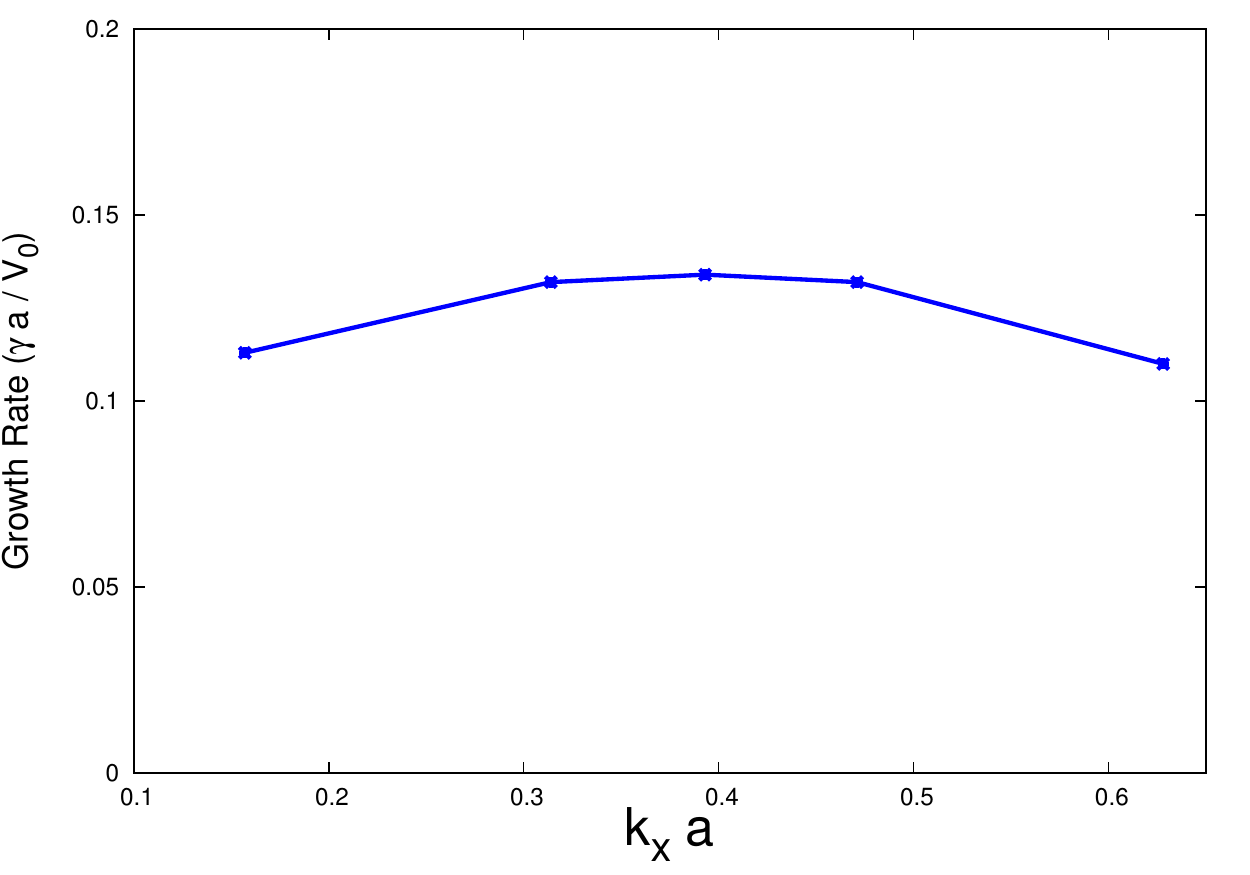}
\caption{(Color online) (a) Time evolution of kinetic energy along the perpendicular to the flow direction is evaluated with time. Fig 2 of Keppens et al \cite{keppens:1999} reproduced by CFD-2D for the identical parameters with Mach Number (M) = 0.5. The value of growth rate ($\gamma$ = 1.7) is found to be in very good agreement with the value obtained from Table 1 ($\gamma$ = 1.728) by Keppens et al \cite{keppens:1999} (b) The growth rate ($\frac{\gamma a}{V_0}$) of K-H instability at $k_x = \{1,2,2.5,3,4\}\pi$ with Mach number $M = 0.5$ identical to Keppens et al \cite{keppens:1999}, where, $V_0$ is the maximum velocity.}
\label{keppens}
\end{center}
\end{figure}

In order to check the correctness of the code in the compressible regime, we reproduce growth rates for the hydrodynamic limit numerically obtained by Keppens et al \cite{keppens:1999} for a compressible K-H flow [Fig \ref{keppens}(a)] at different mode numbers [Fig \ref{keppens}(b)] of excitation. For example, we start with a flow of velocity profile $u_x = U_0 \left[\tanh(\frac{y - L_y/3}{a}) - \tanh(\frac{y - 2 L_y/3}{a}) - 1\right]$ with a shear width $a = 0.05$ and excite a sinusoidal perturbation in perpendicular to the flow velocity $u_y = u_{y_0} \sin (k_x x) \exp\left[{-\frac{(y - L_y/3)^2}{\sigma^2}}\right] + u_{y_0} \sin (k_x x) \exp\left[{-\frac{(y - 2L_y/3)^2}{\sigma^2}}\right]$ with $U_0 = 0.645$, $u_{y_0} = 10^{-4}$, $k_x = \{1,2,2.5,3,4\}\pi$ and $\sigma = 4 a$. We keep $\gamma = \frac{5}{3}$, $P_0 = 1 = \rho_0$, $L_x = 1$ and $L_y = 2$ with $N_x = N_y = 128$ and Mach number $M = 0.5$. We find that the growth rates agree well with the results reported earlier.\\

\subsection{Parameter details}
Having extensively benchmarked the newly developed compressible code CFD-2D, we proceed to simulate the vortex merger problem discussed above. We choose the system size to be big enough so that, the vortex mergers will not get affected by the mergers in the mirror domains due to periodic boundary conditions. We keep the value of $\mu$ small enough such that, we are able to isolate the effect of compressibility. For numerical purpose, we further choose a finite radius of each point vortices as $d_{pv} = 0.032 \frac{L}{2}$. We define the patch turnover time as $\tau_D = \frac{2 \pi}{\omega_{pa}} = \frac{2 \pi}{1} = 2 \pi$. For further check, we compare our results with identical parameter set to AG-Spect \cite{akanksha:2017}. We evolve the prearranged vortices in a system of area $(4\pi)^2$ with a resolution of $256^2$ grids with time-steps of $10^{-5}$. We obtain the identical result with another run with grid size $512^2$. This indicates the numerical convergence of our results. For the rest of the paper, unless otherwise indicated, we keep the grid resolution at $256^2$. The parameter regime for which we compare the results of the two codes (CFD-2D and AG-Spect) are given in the following table (Table \ref{parameter}).\\
\begin{table}[h!]
\centering
\begin{tabular}{ |c|c|c|c|c|c|c|c|c| }
 \hline
 $N_x = N_y$ & $L_x = L_y$ & $dt$ & $\mu$ & $R_{pa}$ & $R_{pv}$ & $N_{pv}$ & $\omega_{pa}$ & $\omega_{pv}$ \\
 \hline
 ~ & ~ & ~ & ~ & ~ & ~ & ~ & ~ &\\
 $256$ & $4 \pi$ & $10^{-5}$ & $2 \times 10^{-4}$ & $0.3\frac{L}{2}$ & $0.4 \frac{L}{2}$ & $5$ & 1 & $10$ \\
 ~ & ~ & ~ & ~ & ~ & ~ & ~ & ~ &\\ 
 \hline
\end{tabular}
\caption{Parameter details for the benchmark of the code CFD-2D with AG-Spect. These parameters are kept identical throughout this report unless stated otherwise.}
\label{parameter}
\end{table}

\begin{figure}[h!]
\begin{center}
\includegraphics[scale=0.65]{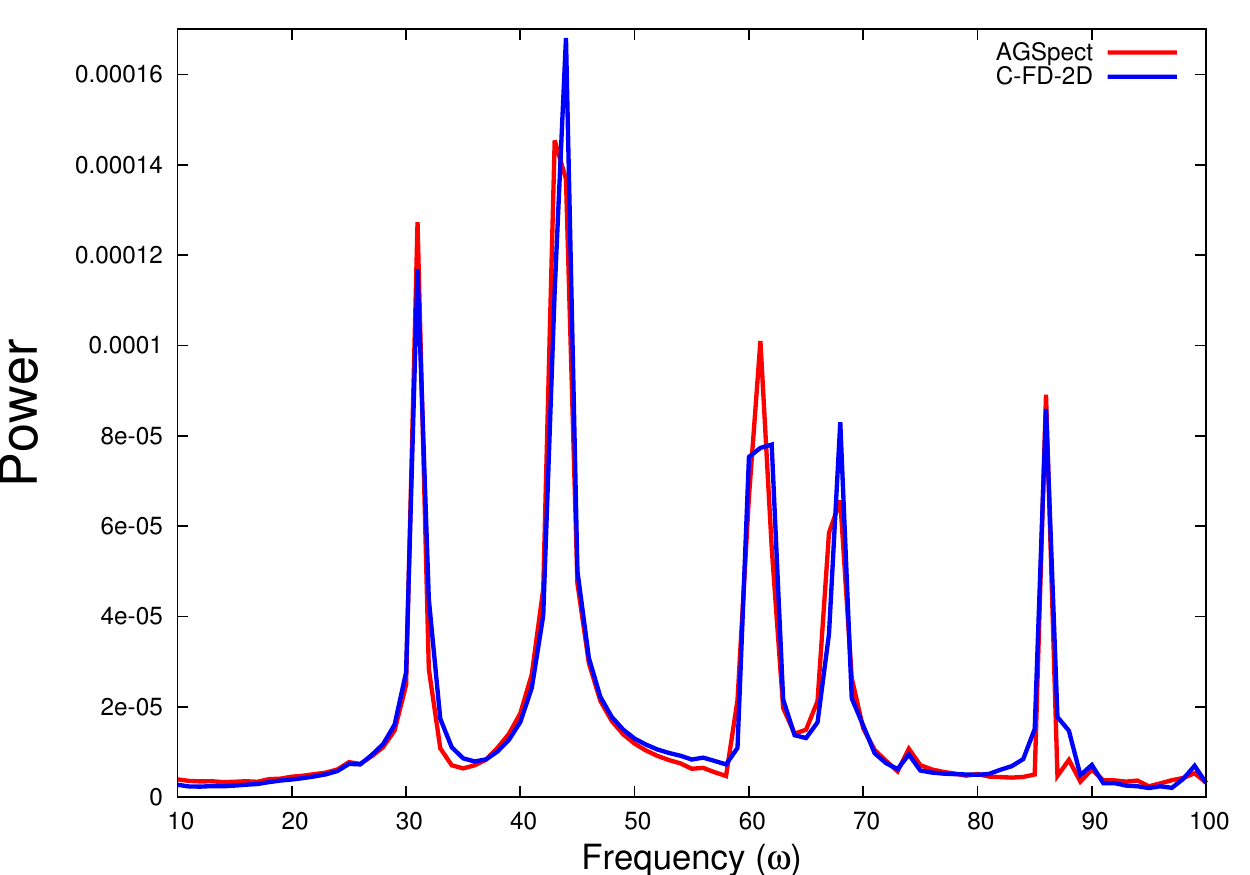}
\caption{(Color online) Comparison of natural frequency and its harmonics as well as its beats with its harmonics from two different codes, AGSpect\cite{akanksha:2017} and CFD-2D (this code) for $M = 0.5$}
\label{comparison}
\end{center}
\end{figure}

We find the results obtained from the two independent codes in both the incompressible ($M = 0.05$) and the  compressible ($M = 0.5$) limits, to be quite satisfactory [Fig \ref{comparison}].\\

We keep the above set of parameters identical throughout our paper unless stated otherwise. Next we vary only compressibility via Mach Number ($M$) from $0.05$ to $0.8$ and make the following observations.


\section{Results}

\subsection{Results in incompressible limit}
For the incompressible limit, both the codes (CFD-2D and AG-Spect) are found to replicate the results earlier obtained \cite{ganesh:2002} through electrostatic particle-in-cell (PIC) method. Also the results qualitatively match with the recent observation by Swaminathan et al \cite{perlekar:2016} addressed using Fourier pseudo-spectral technique.  A detailed and complete description of the merging process is depicted in these \cite{ganesh:2002,perlekar:2016} earlier communications in incompressible limit ($M \equiv 0$).\\

\begin{figure*}[t]
\begin{center}
\includegraphics[scale=0.45]{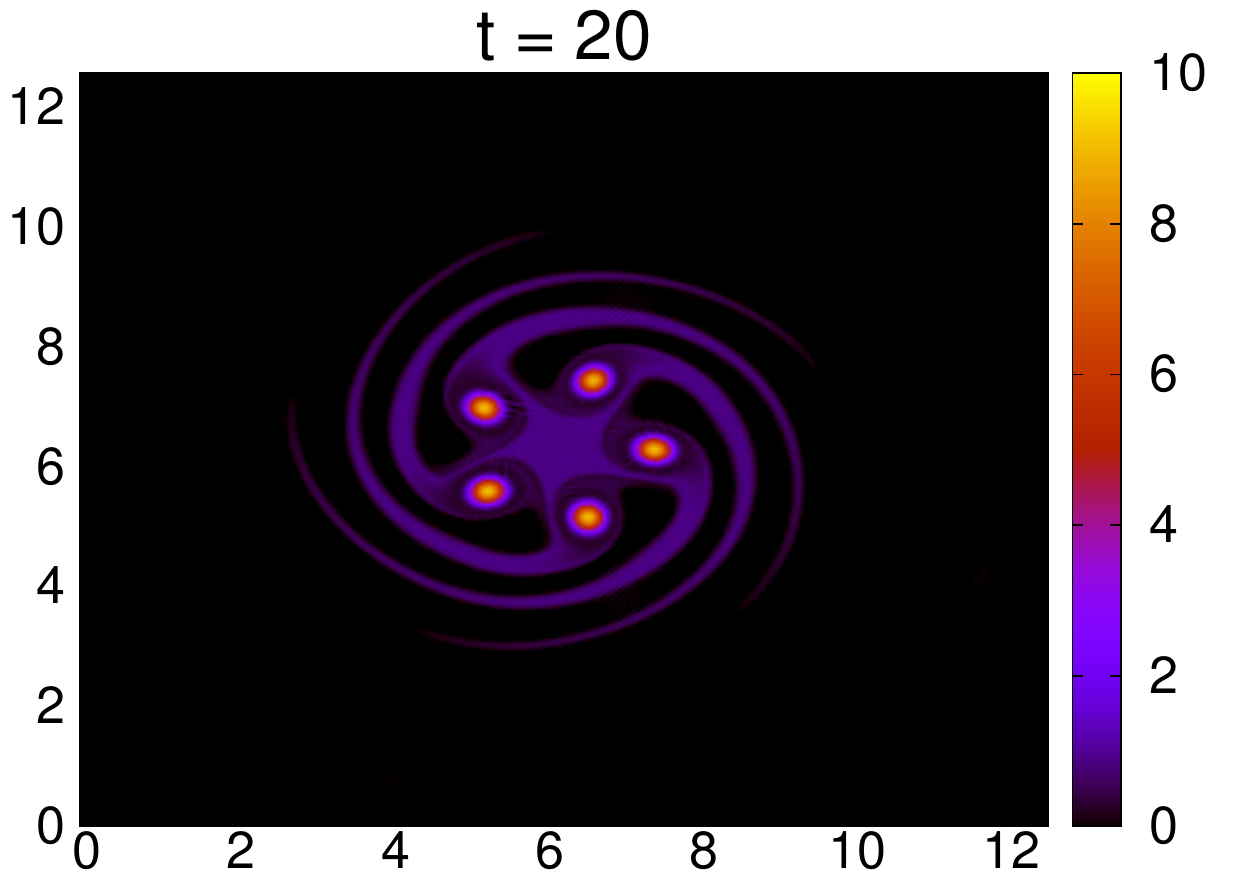}
\includegraphics[scale=0.45]{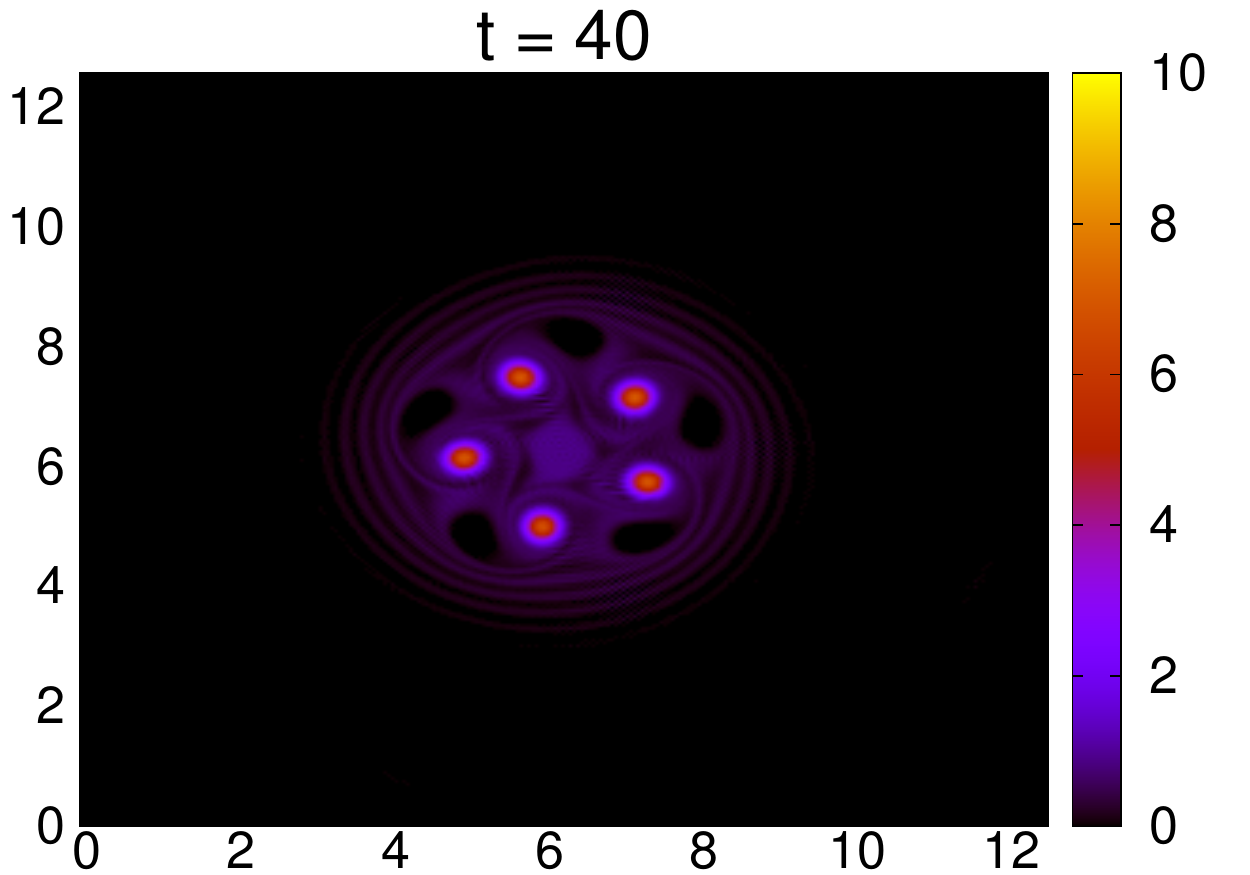}
\includegraphics[scale=0.45]{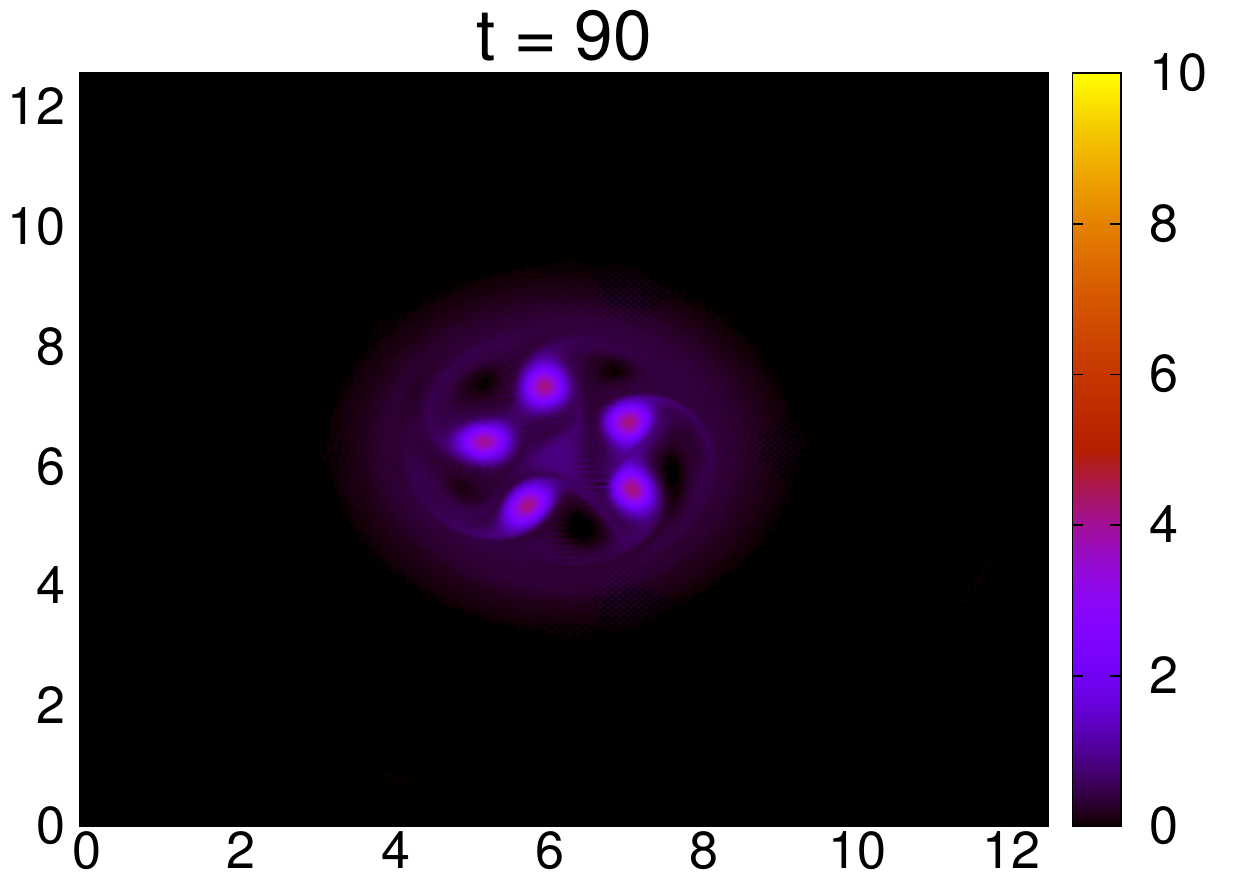}
\caption{(Color online) Time evolution of the prearranged vortex merger with $N_{pv} = 5$ and $M = 0.05$ (incompressible) with grid size $256^2$ with the parameters mentioned in Table \ref{parameter}.}
\label{snap_incompressible}
\end{center}
\end{figure*}

As found earlier, we find the Rankine vortex generates V-states and undergoes filamentation process. The point-like vortices become unstable and start falling towards the central vortex. The filamented fingers wave-break and  engulf the point-like vortices. Then the vortex-hole pattern generates and remains quasistationary for a substantially long time scale (at least upto $t = 90 \simeq 14 \tau_D$, where, $\tau_D$ is the vortex patch turn-over time) forming a \textquotedblleft vortex crystal\textquotedblright. In an unscreened or scale-less hydrodynamics it was shown that such stationary crystals are possible \cite{lansky:1997}. The snapshots of this incompressible ($M = 0.05$) vortex merger system are shown in Fig \ref{snap_incompressible}.\\

\subsection{Results in compressible limit}
\begin{figure}[h!]
\begin{center}
\includegraphics[scale=0.65]{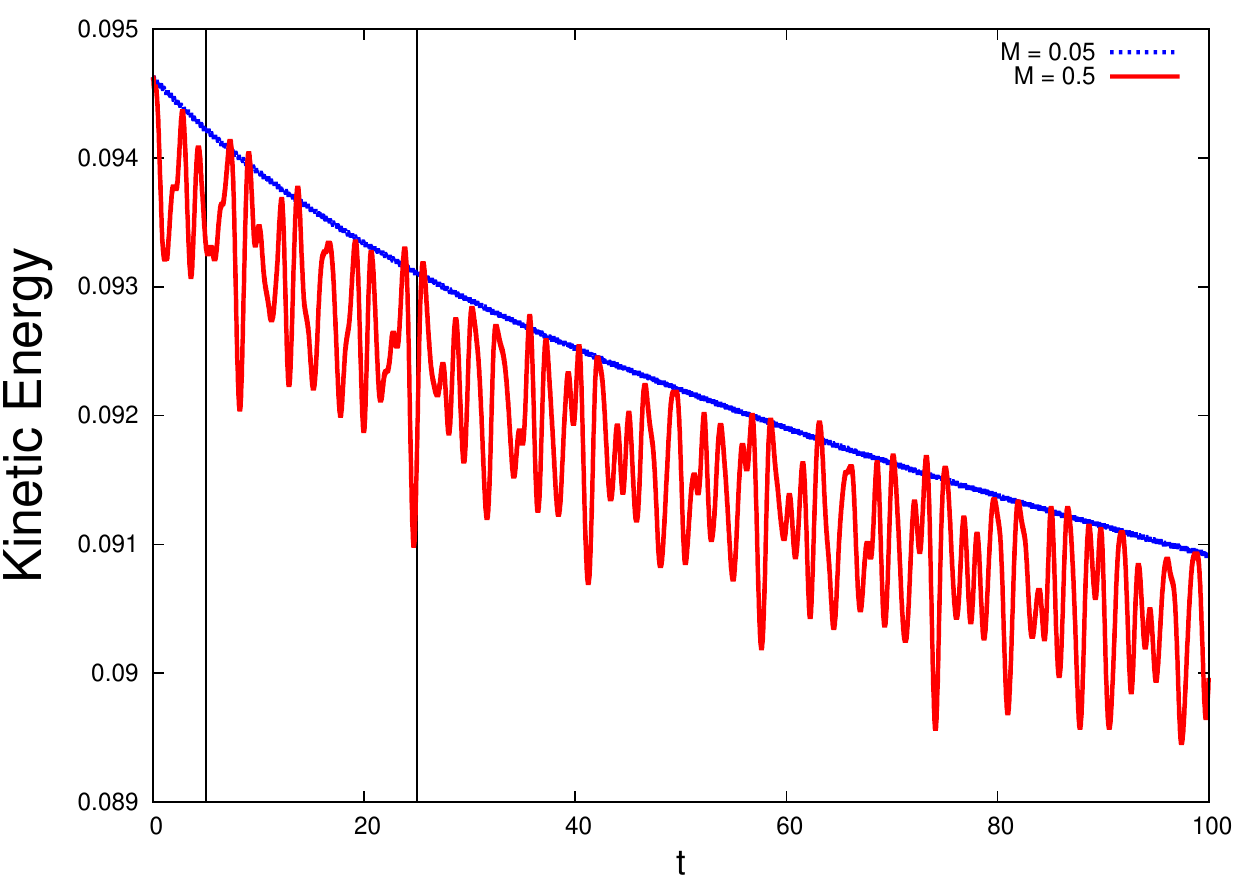}
\caption{(Color online) The red solid line represents the time evolution of total kinetic energy for M = 0.5 while the blue dotted line represents the same for M = 0.05. The decay in the total energy is due to the small viscosity $\nu = 2 \times 10^{-4}$ and follows the analytical profile $\frac{U_0^2}{2} e^{-2\nu t}$. The vertical black lines at $t = 5$ and $25$ represents the start and end of merging process respectively at M = 0.5.}
\label{KE}
\end{center}
\end{figure}

As the Mach number ($M$) is increased, the qualitative behavior of the merging of the vortices and the quasistationary vortex and transient hole patterns remain the same. Also, the total kinetic energy starts oscillating with time (Fig.\ref{KE}).The next section is devoted to analysing the frequency of oscillation, obtaining an order of estimate of the fundamental frequency from a theoretical model and the scaling of these frequencies with Mach number. \\

The merging time is found to get affected with the increase in compressibility. In the table below, we present the merging time with different Mach number ($M$). $T_s$ represents the time after which the Kelvin waves from the surface of the patch vortex touch the point like vortices, $T_e$ represents the time when the holes (region of zero vorticity) gets isolated and fully trapped within distorted patch vortex and $T_m = T_e - T_s$ represents the time difference between these two processes. $T_d$ represents the time after which the quasistationary vortex crystal gets destructed due to nonlinear compressibility effects.

\begin{table}[h!]
\centering
 \begin{tabular}{ |c|c|c|c|c|c|c|c|c|c| }
 \hline
 $M$ & 0.05 & 0.1 & 0.2 & 0.3 & 0.4 & 0.5 & 0.6 & 0.7 & 0.8 \\
 \hline
 $T_s$ & 4.8 & 4.8 & 4.8 & 4.9 & 4.9 & 5.0 & 5.0 & 5.1 & 5.1 \\
 \hline
 $T_e$ & 24.1 & 24.2 & 24.3 & 24.5 & 24.7 & 25.0 & 25.3 & 25.8 & 26.2 \\
 \hline
 $T_m$ & 19.3 & 19.4 & 19.5 & 19.6 & 19.8 & 20.0 & 20.3 & 20.7 & 21.1 \\
 \hline
 $T_d$ & 90.0 & 87.5 & 85.0 & 79.0 & 77.5 & 75.5 & 74.5 & 72.5 & 72.0\\ 
 \hline
 \end{tabular}
\caption{Merging time and vortex crystal destruction time with different compressibility for $N_{pv} = 5$.}
\label{merging_time}
\end{table}

From the above table, it is found that the startup of the merging process ($T_s$) gets delayed and the merging time ($T_m$) also gets elongated as the compressibility is increased. We have attempted to give a theoretical estimate of these numbers from a simple analytical model in the next section.\\

\begin{figure*}[t]
\begin{center}
\includegraphics[scale=0.45]{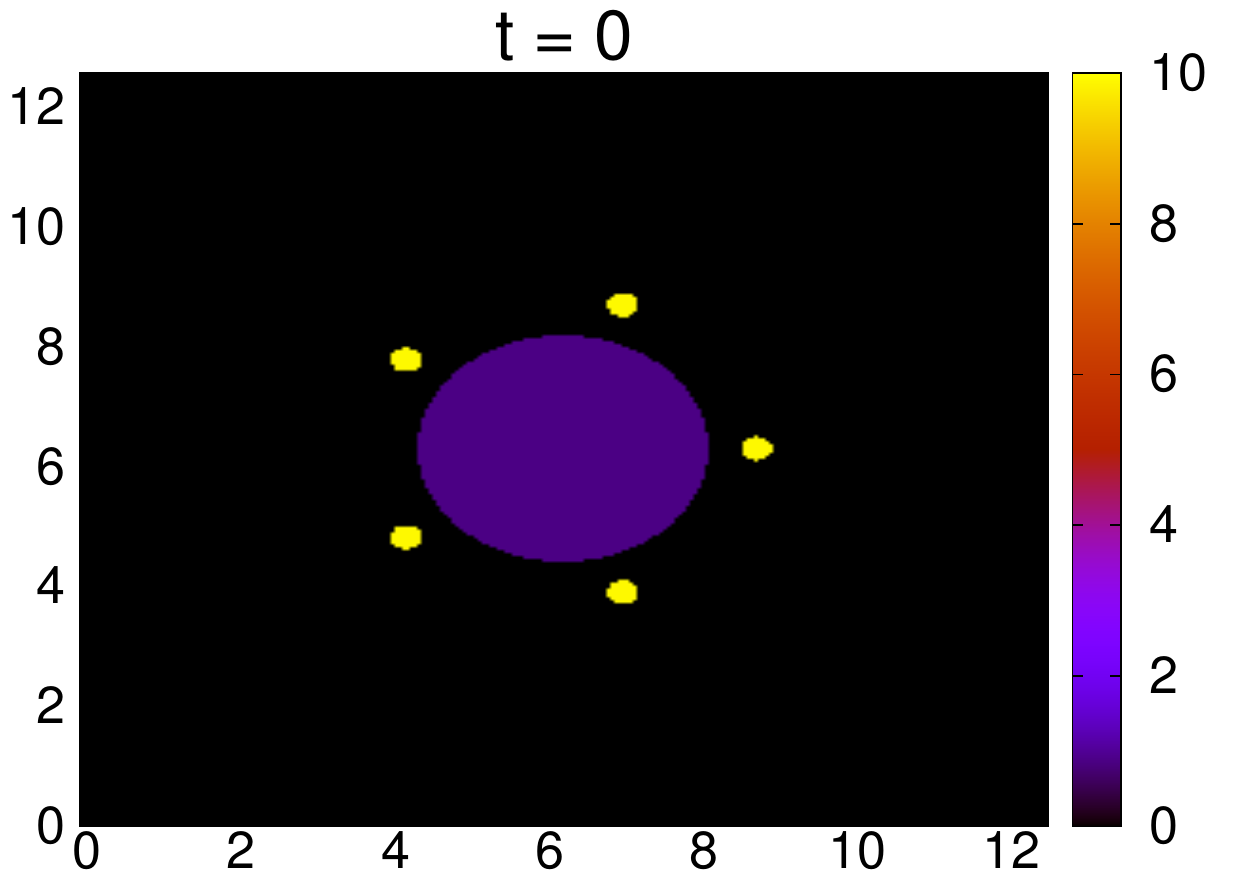}
\includegraphics[scale=0.45]{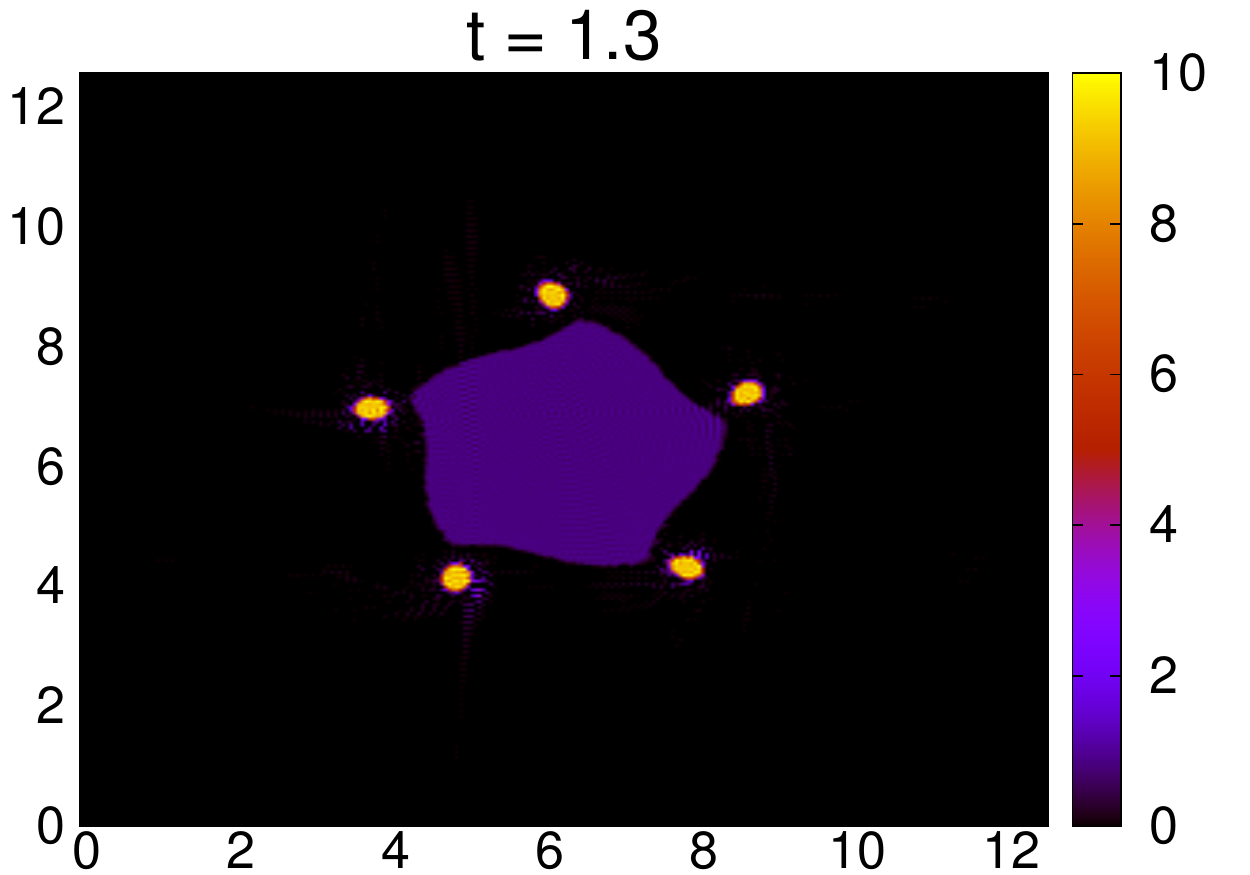}
\includegraphics[scale=0.45]{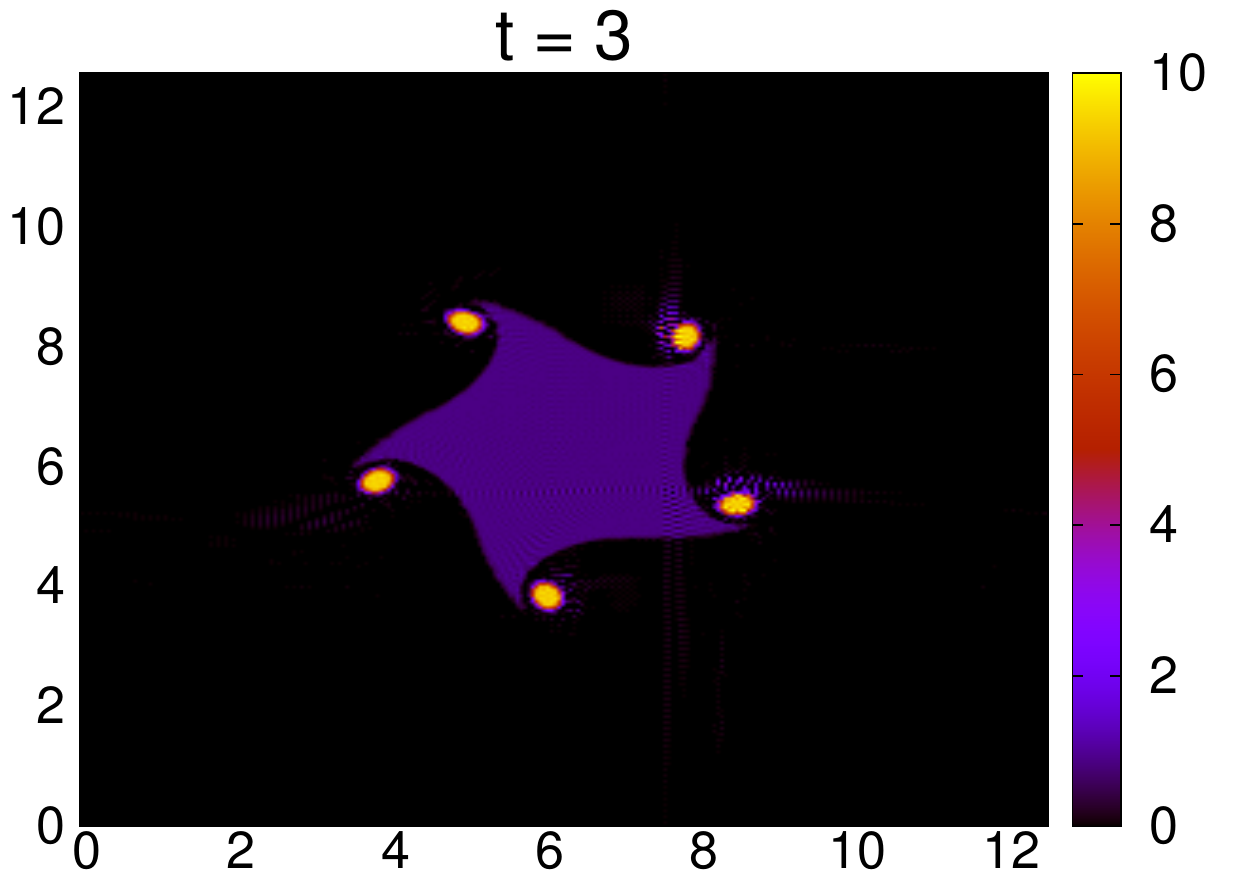}\\

\includegraphics[scale=0.45]{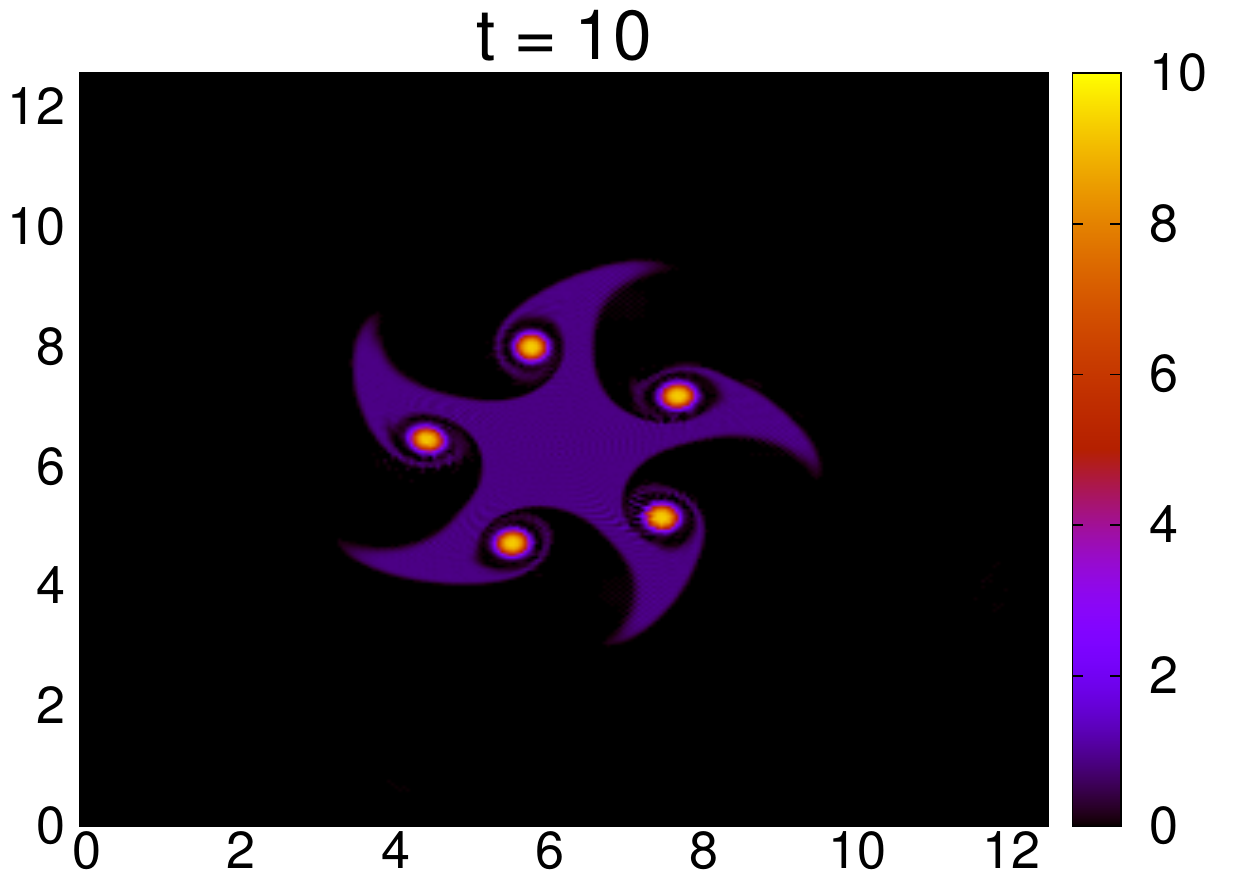}
\includegraphics[scale=0.45]{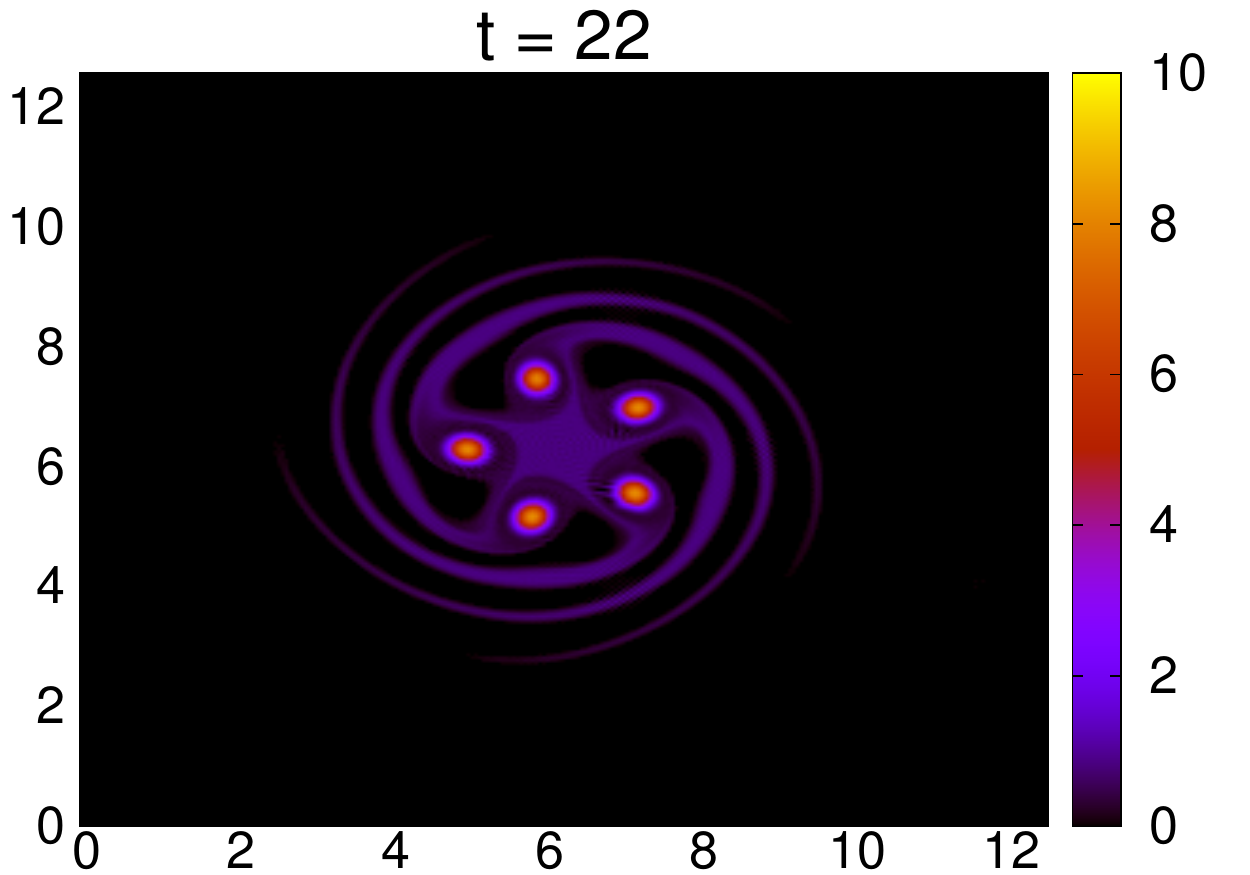}
\includegraphics[scale=0.45]{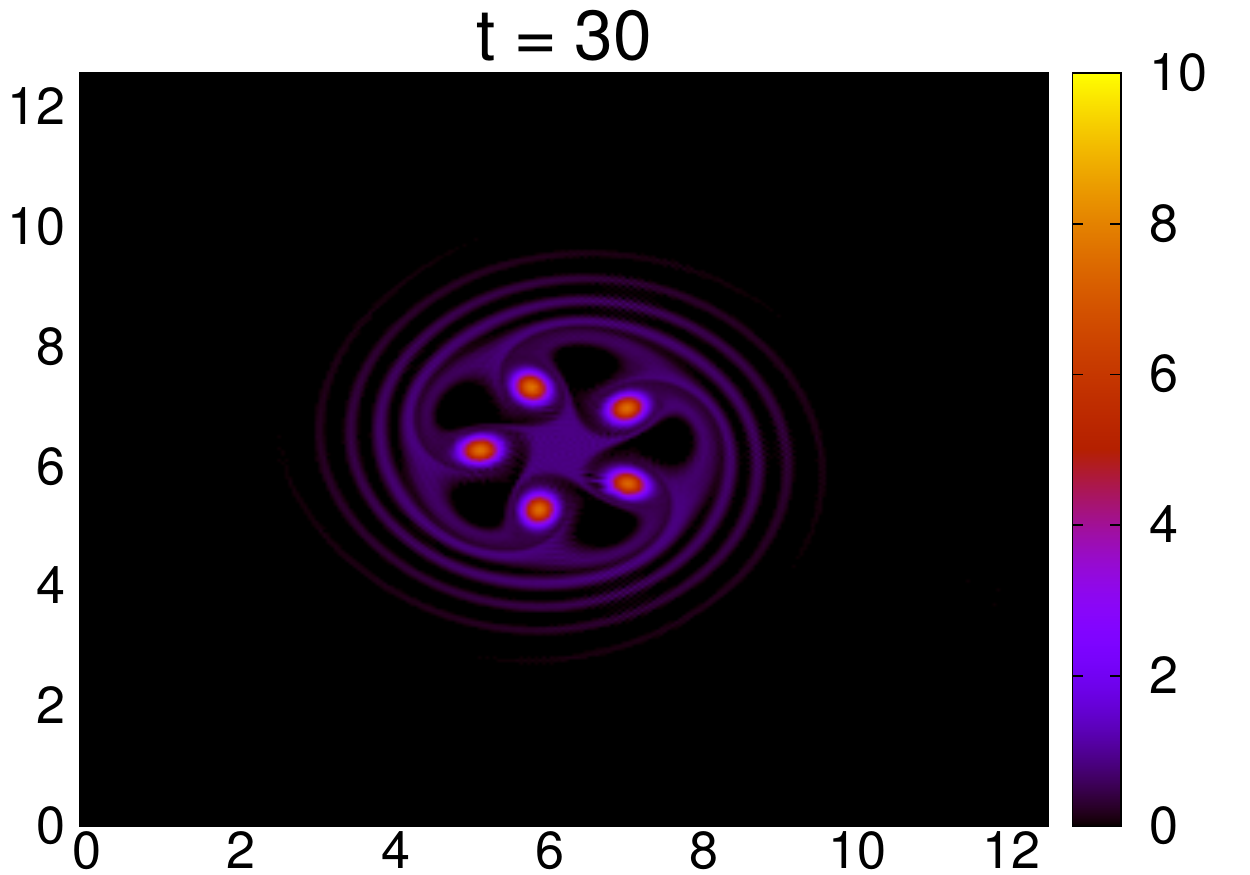}\\

\includegraphics[scale=0.45]{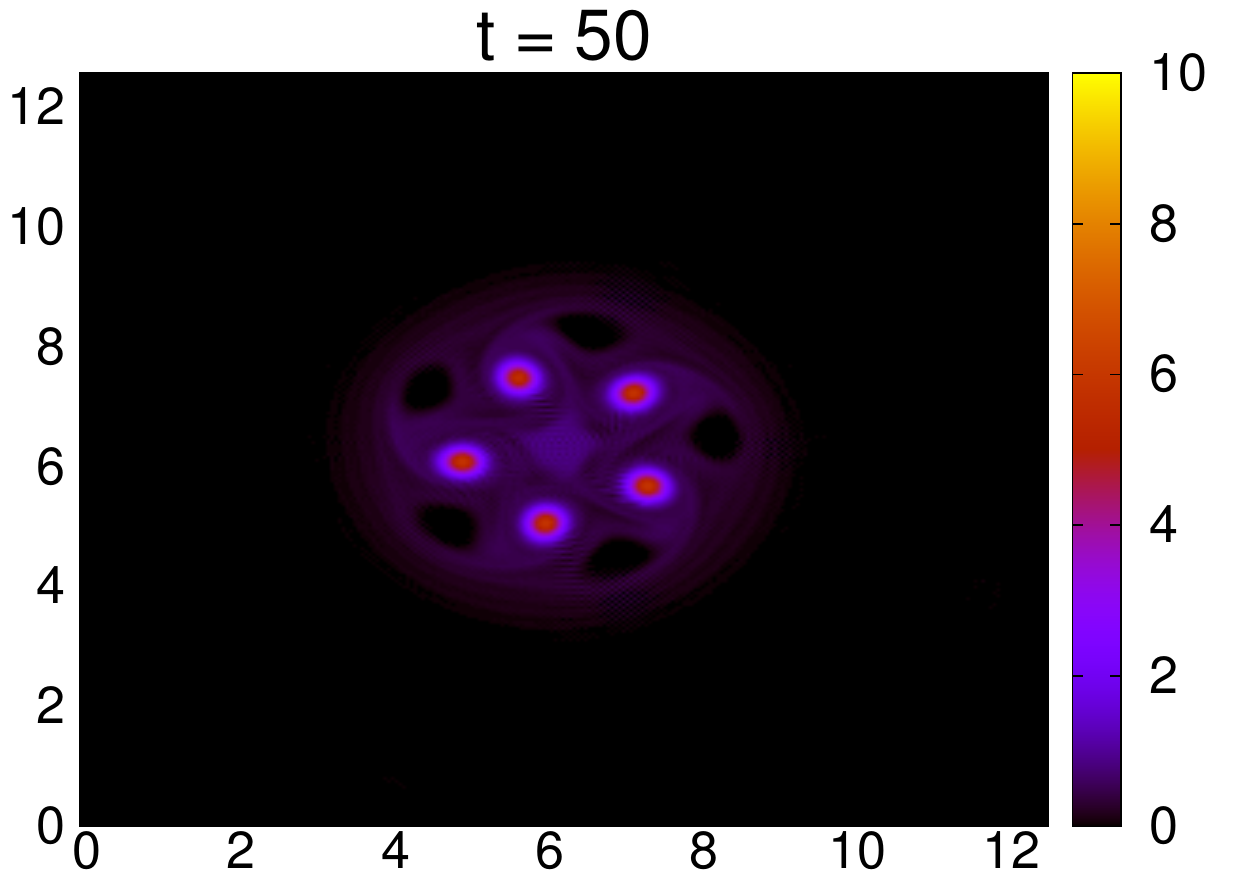}
\includegraphics[scale=0.45]{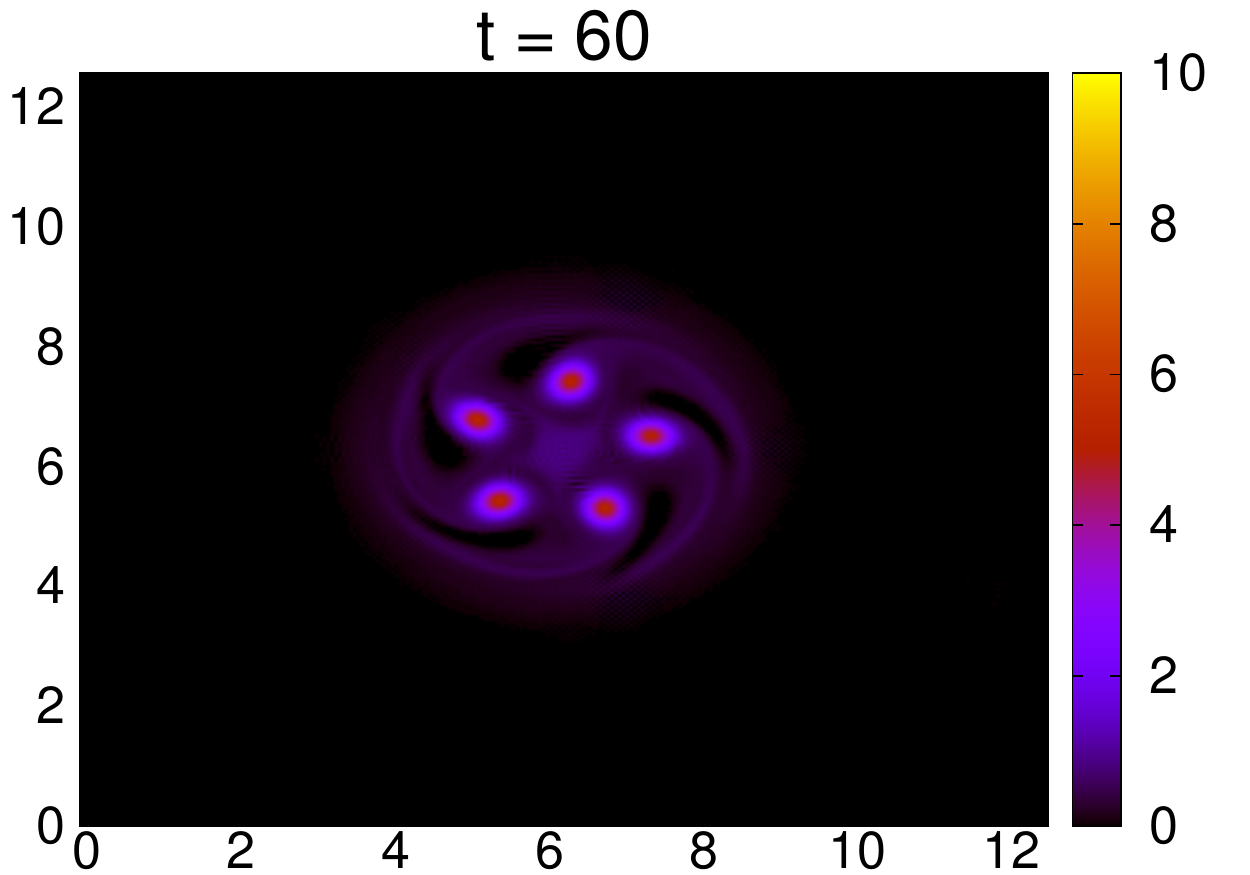}
\includegraphics[scale=0.45]{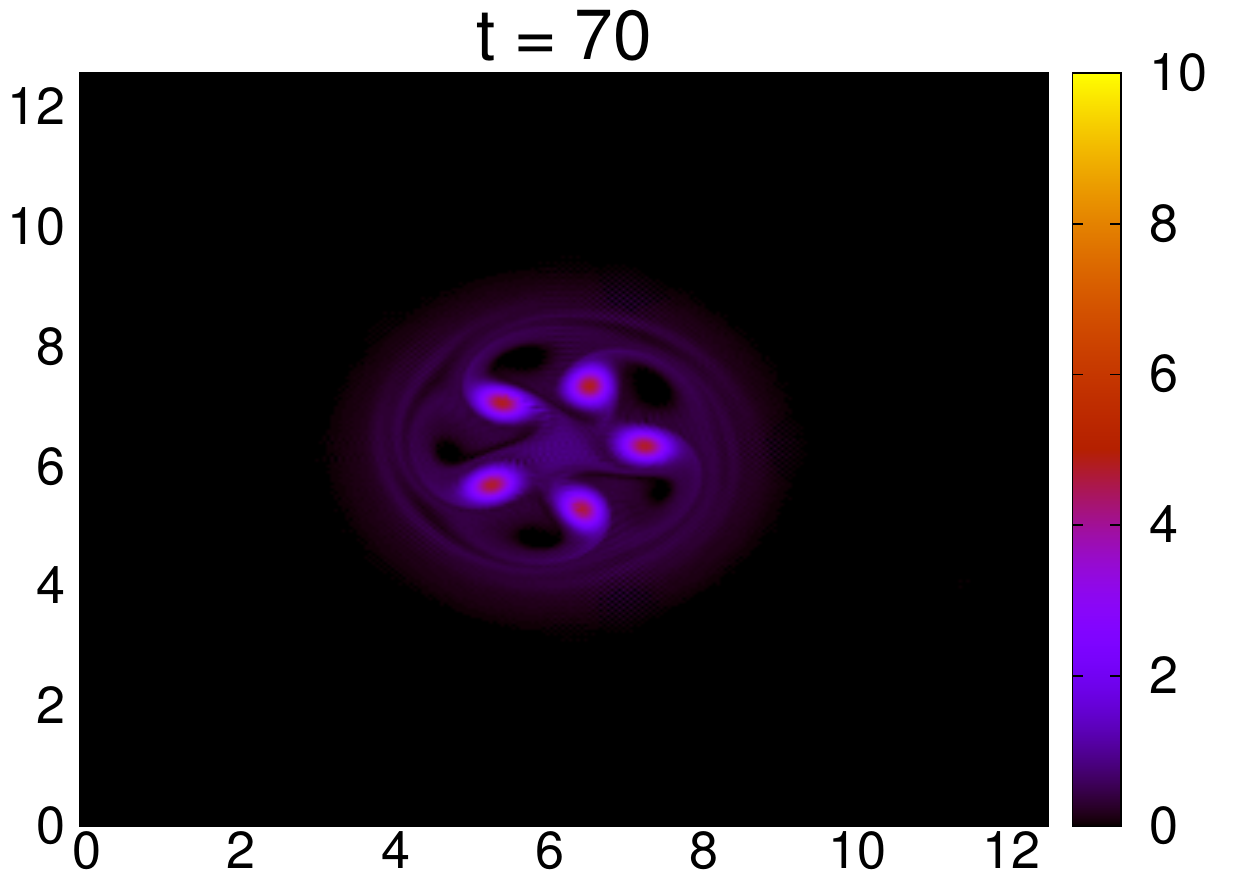}\\

\includegraphics[scale=0.45]{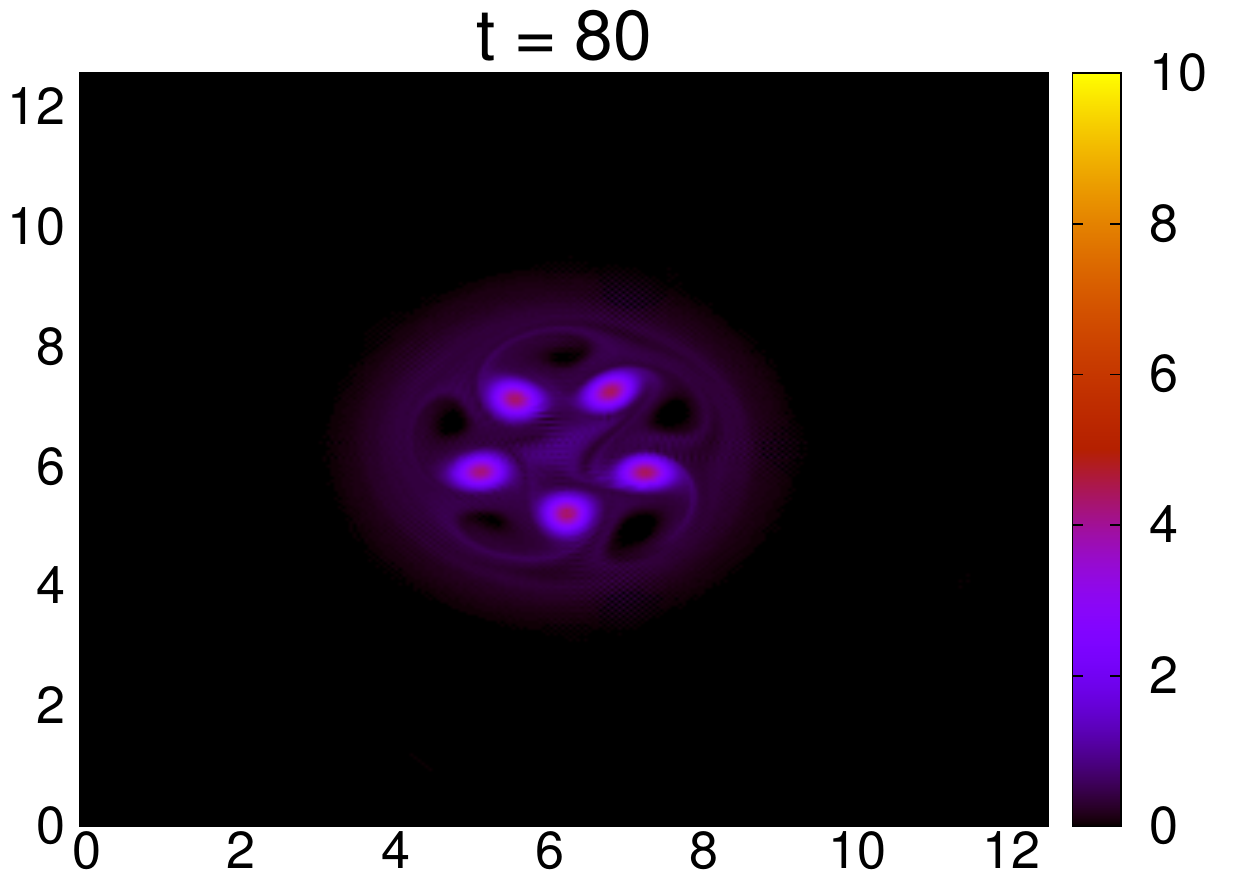}
\includegraphics[scale=0.45]{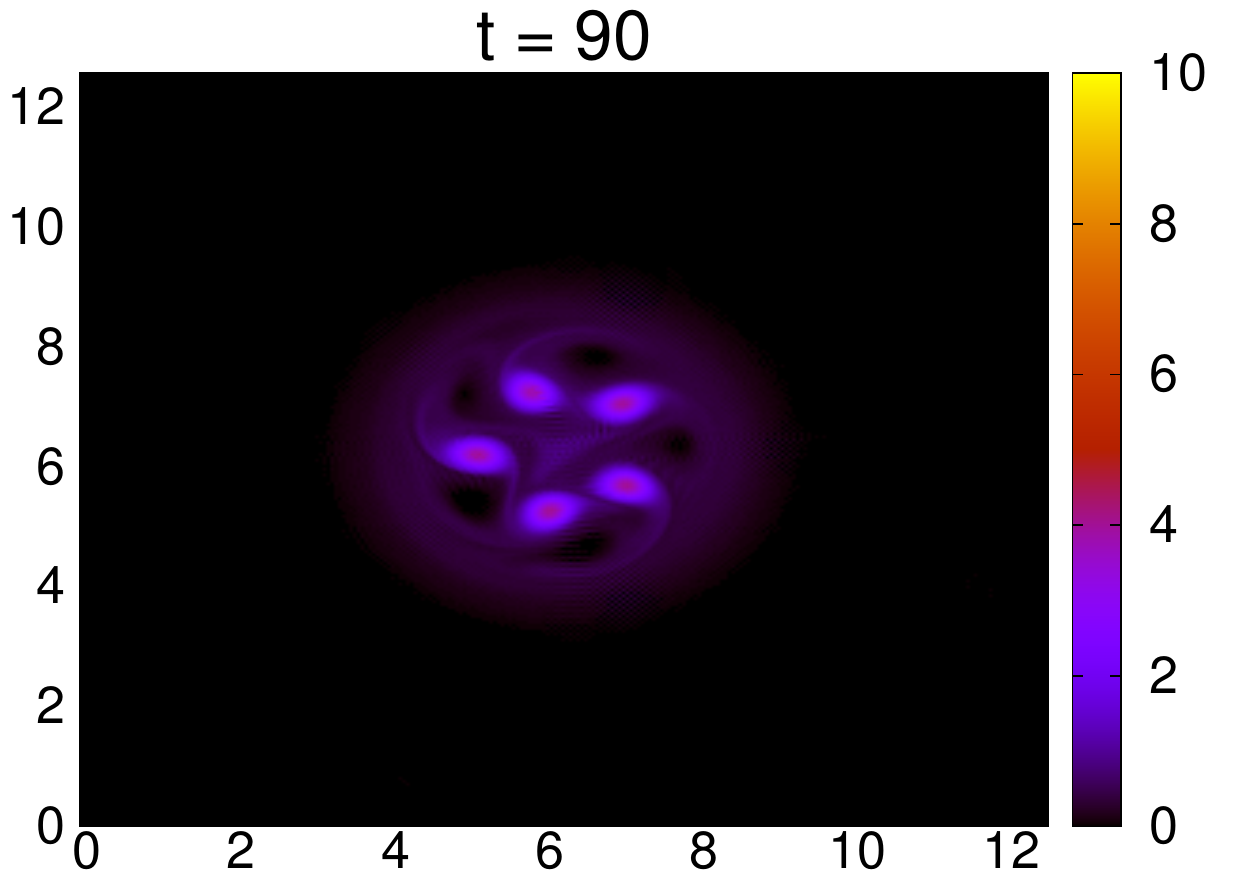}
\includegraphics[scale=0.45]{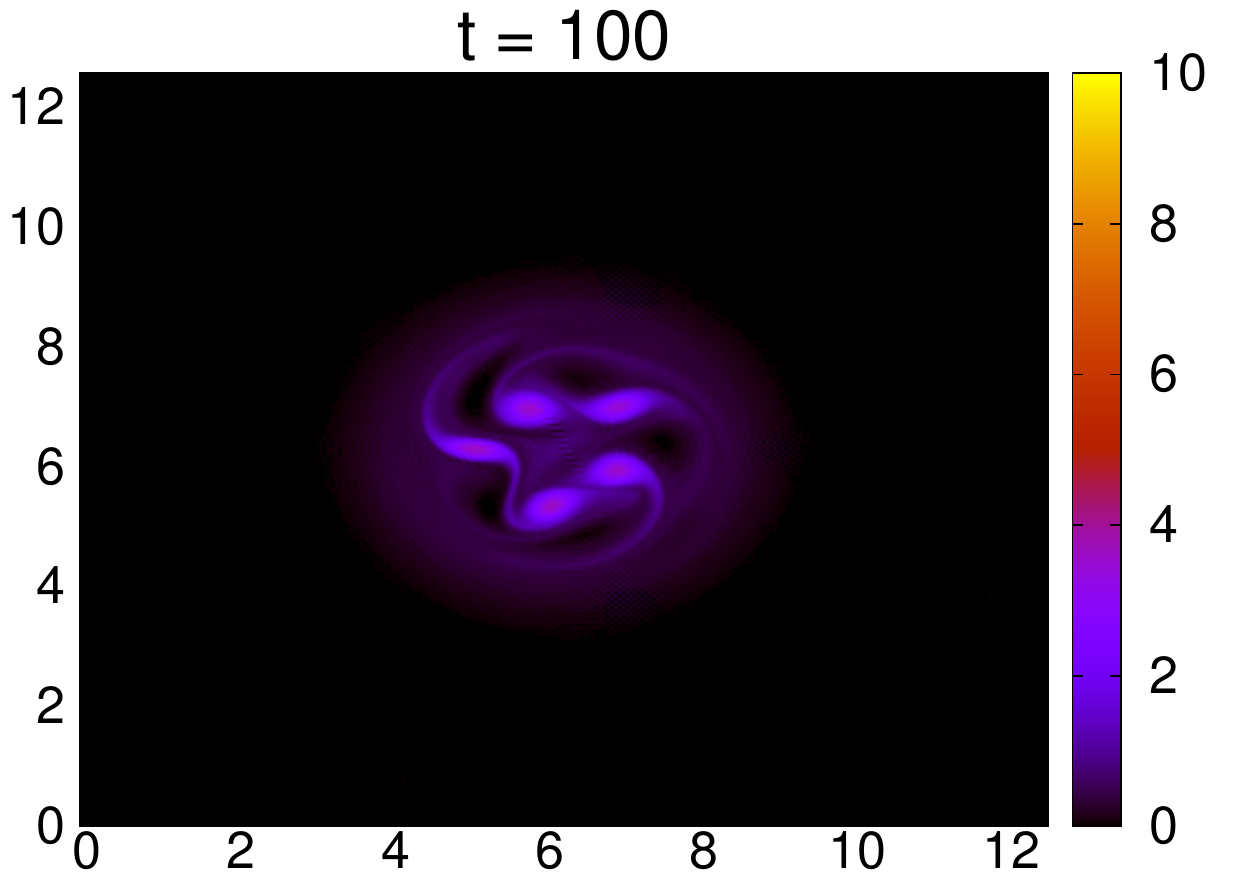}
\caption{(Color online) Time evolution of the prearranged vortex merger with $N_{pv} = 5$ and $M = 0.5$ for grid size $256^2$ and the parameters mentioned in Table \ref{parameter}.}
\label{snap}
\end{center}
\end{figure*}

Fig \ref{snap} represents selected snapshots of the vortex mergers and quasistationary vortex and transient hole patterns for M = 0.5. From the last snapshot at time $t = 100 \simeq 16 \tau_D$, we have found that the vortex crystal pattern gets distorted and eventually in a long run the vortices merge and forms a single vortex. As the compressibility is increased the life-time of the vortex crystal is found to decay. The time after which the vortex crystal gets distorted ($T_d$) is evaluated with \textquotedblleft eye\textquotedblright ~ estimation and has been summarised in Table \ref{merging_time} for different Mach number ($M$). The sonic waves generated due to compressibility, perturb the crystal structure continuously and cause the melting. A detailed discussion is provided in the next section.\\

\section{Analysis of the Results for $N_{pv} = 5$}
For the incompressible limit ($M = 0$), we calculate the analytical decay rate of the total kinetic energy with time due to viscous dissipation. The decay is found to be in good agreement with the numerically obtained data with $M = 0$. Next we detrend the decay from the total kinetic energy of each of the runs with different Mach numbers. Then with the \textquotedblleft detrened kinetic energy\textquotedblright~ we determine the frequency of the time series data. For the frequency calculation we use the standard FFTW libraries (v-3.3.3) \cite{FFTW3:2005}.\\

In the Fourier transform of the full kinetic energy data, we find five distinct peaks in the power spectrum. We redo the similar calculation for Mach numbers (M) = 0.1, 0.2, 0.3, 0.4, 0.5, 0.6, 0.7, 0.8. We believe the first peak corresponds to the natural frequency of the compressible system occurring due to the interaction between the point like vortices as well as the central patch vortex. The third  peak has the frequency twice of the first mode. The second is the beat frequency of the first and third mode. The frequency of the fifth mode is the double of the second mode. The fourth is the beat frequency of the second and fifth mode. If we represent the five prominent peaks in the power spectrum A, B, C, D, E in the increasing order of the frequency, we observe, the following relaton as stated above:\\

$A; ~ B = \frac{A+C}{2}; ~ C = 2A; ~ D = \frac{B+E}{2}; ~ E = 2B$\\

\noindent
We find this relation to be true for all the values of M mentioned above.\\

\begin{figure}[h!]
\begin{center}
\includegraphics[scale=0.65]{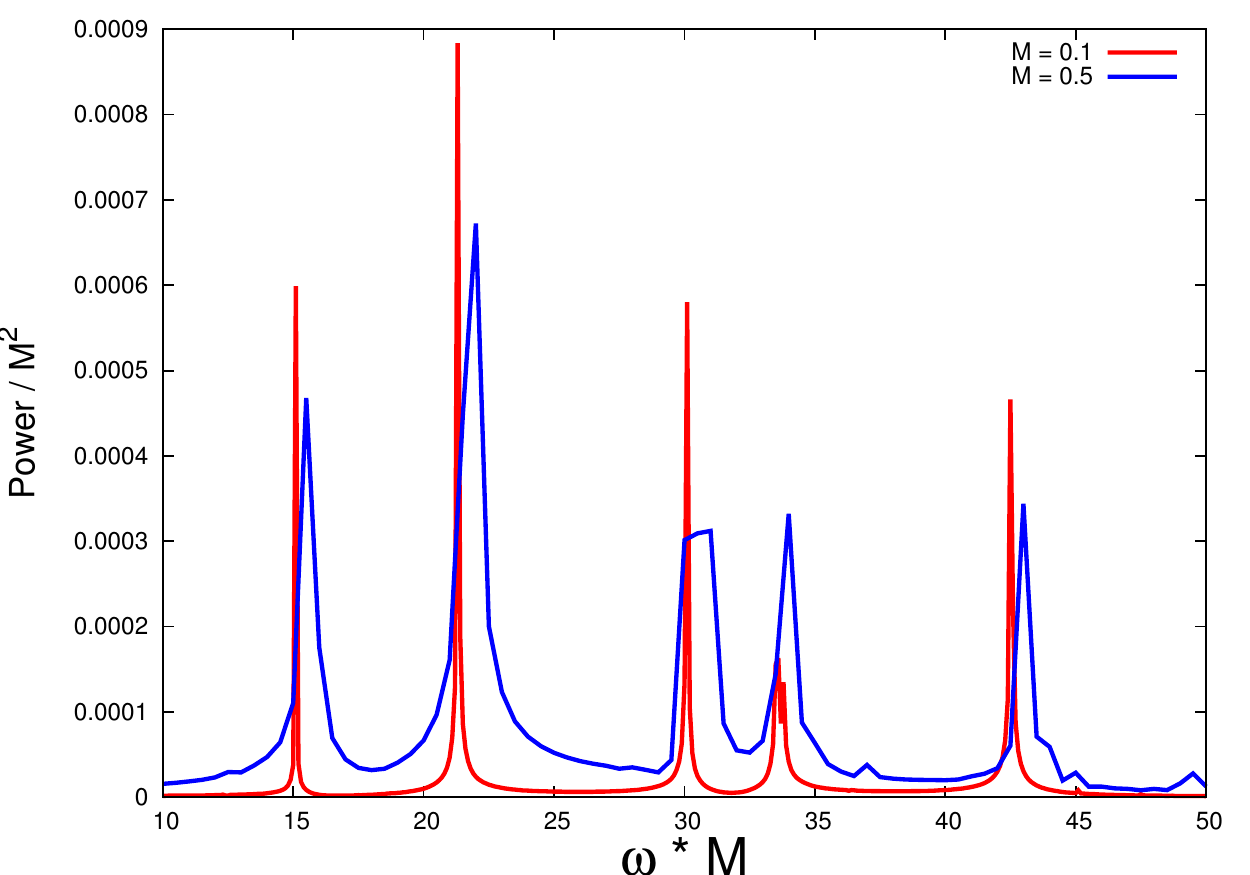}
\caption{(Color online) The red solid line represents the freqencies present in total kinetic energy for M = 0.1 while the blue dotted line represents the same for M = 0.5. The power of kinetic energy is found to vary as $M^{-2}$. }
\label{freq}
\end{center}
\end{figure}

Next we turn to the scaling of the frequency of the fundamental as well as all the other modes. From the basic principle of oscillations\cite{moore} in a compressible fluid, we know that the $\omega_0 \propto \frac{1}{l}\frac{\gamma P}{\rho}$, where, $\omega$ is the natural frequency of the fluid and $l$ is some lengthscale defined later. Thus $\omega_0 \propto C_s^2 = \frac{U_0}{M}$, where, $U_0$ is the maximum velocity of the fluid. Hence we see, $\omega_0 \cdot M = $ constant. We have plotted all the frequencies obtained from the fourier transform of the total kinetic energy data and plotted in a rescaled axis of $\omega \cdot M$ and find every peak with different compressibility falls on top of the respective ones as shown in Fig \ref{freq}. \\

\begin{figure}[h!]
\begin{center}
\includegraphics[scale=0.65]{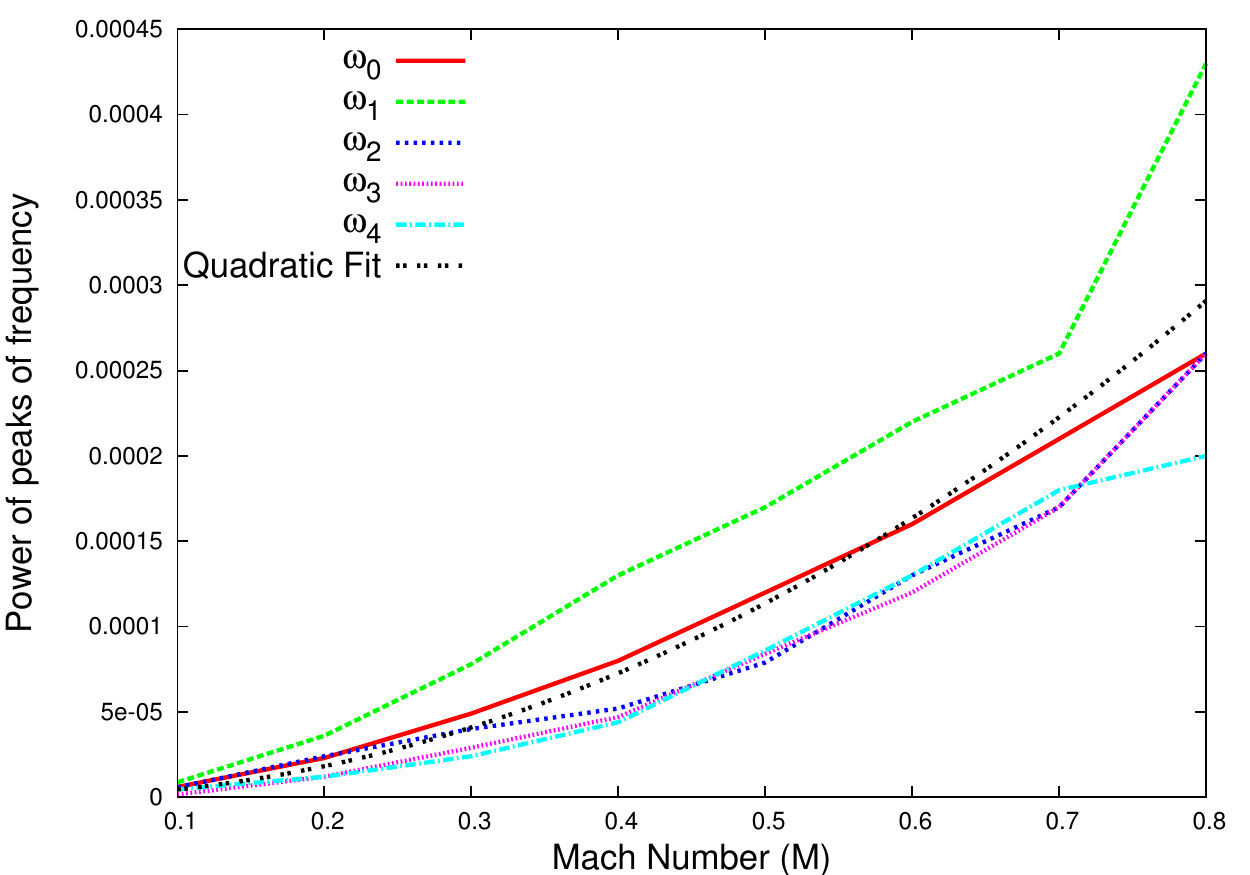}
\caption{(Color online) The red solid line represents the power of natural frequency of kinetic energy for different $M$. The green dashed ($\omega_1$), blue dotted ($\omega_2$), magenta dashed-dotted ($\omega_3$) and the sky long dashed ($\omega_4$) lines represents the same for second, third, fourth, fifth peaks respectively. The black double-dotted line is a quadratic fit for the fundamental frequency $\omega_0$ i.e. the red solid line.}
\label{power_data}
\end{center}
\end{figure}

It is well-known that \cite{moore} the amplitude of sound wave in a compressible fluid within a pipe is given by $A = C_s^2 \rho k \xi_0$ where $k$ is the wave vector and $\xi_0$ is the amplitude of displacement. Hence, $A \propto C_s^2 = \frac{U_0^2}{M^2}$ i.e. $A \propto M^{-2}$. Thus the amplitude of sound wave is proportional to the inverse square of Mach number ($M$). Similar reltationship can be found from Fig.\ref{power_data},  where we see that the power of kinetic energy scales as $M^{-2}$. Also with the increase of $M$, from Fig \ref{freq}, we find that consistently the delta-function-like peaks in the fourier spectra gets broadened. The power of kinetic energy in the modes away from the peaks also rises consistently with the increase of $M$.\\

To further understand the role of compressibility on the merger process we divided the full time series data of kinetic energy into three parts viz. a) before merging [$t \leq T_s$], b) upto merging [$t \leq T_e$] and c) after merging [$t \geq T_e$]. We found a fundamental mode and one of its harmonics get excited before the merging process starts. As the merging process proceeds, the relative distances between the vortices decrease nonlinearly resulting in the increase of the frequency of the fundamental mode. Below in Fig \ref{freq_evolution} we show the evolution of fundamental mode as well as its harmonics and beats at the three different time scales.\\

\begin{figure}[h!]
\begin{center}
\includegraphics[scale=0.65]{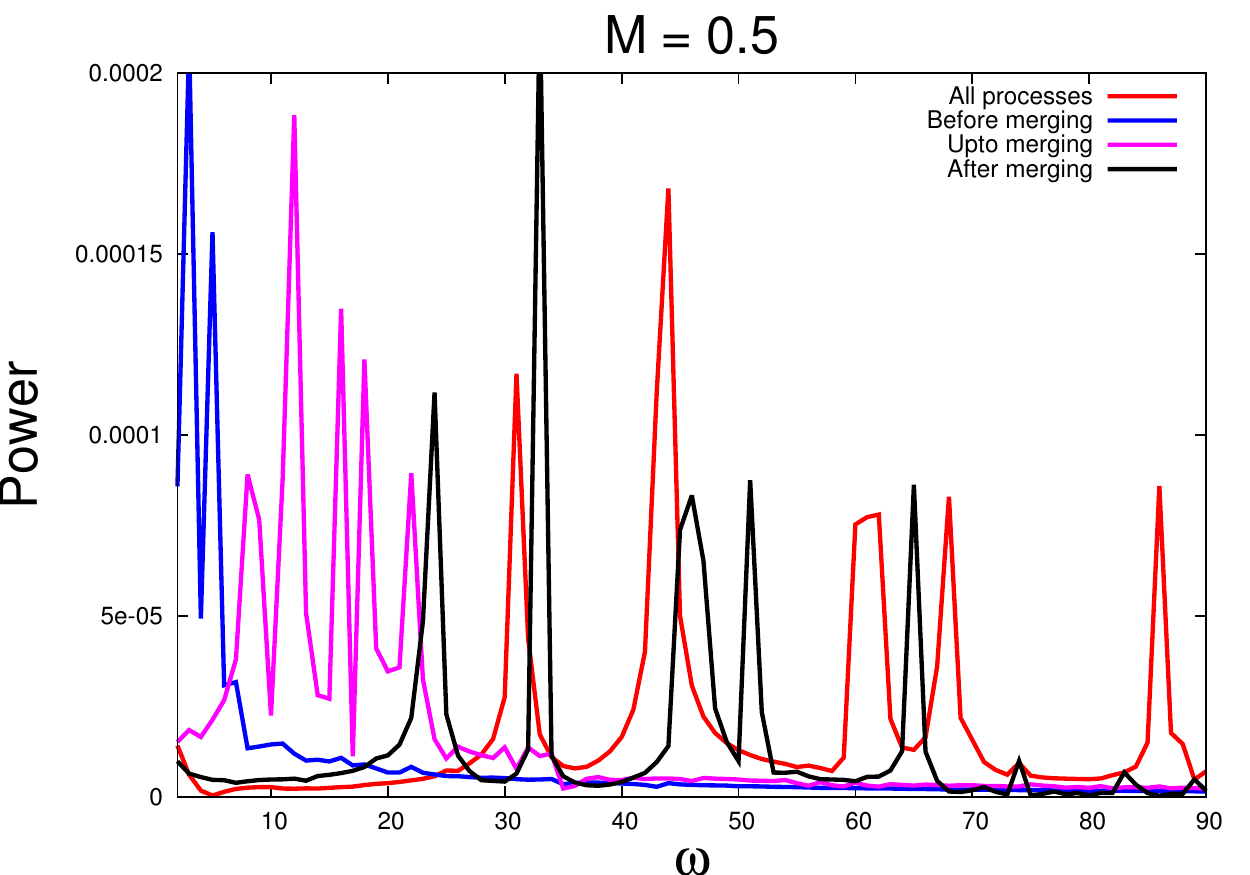}
\caption{(Color online) The frequency of the oscillation is plotted for M = 0.5 at three different timescales. The red solid line indicates the frequencies present in the full evolution. The blue dashed line represents the frequencies present before merging [$t \leq T_s$], magenta dotted line represents the same upto merging [$t \leq T_e$] while the black line is form the data after merging [$t \geq T_e$].}
\label{freq_evolution}
\end{center}
\end{figure}

We attempt an order of estimate of the fundamental frequency at the beginning (the case (a) i.e. before merging [$t \leq T_e$]). We assume that the patch vortices are kept far apart such that at least at the beginning of merging process, they do not interact between each other. Hence we can concentrate on the patch vortex and any one of the identical point like vortices at least before the merging starts. Hence essentially we look into an one dimensional compressible fluid system, whose one end is the surface of the patch vortex and the other end is the surface of the point vortex. We know for a compressible fluid in a long tube the fundamental frequency becomes $\omega_0 = \frac{1}{4l}C_s = \frac{1}{4l} \frac{U_0}{M}$ where $l$ is the length of the tube. The value of $U_0$ can be found from the initial kinetic energy as given in Fig (\ref{KE}). Also the distance between surface of patch vortex and that of point vortex is $d_{patch-point} = (R_{pv} - d_{pv})-R_{pa} = (0.4-0.032) - 0.3 = 0.068 = l$. Hence,
\begin{eqnarray*}
\omega_0 = \frac{1}{4l} \frac{U_0}{M} = \frac{1}{4\times 0.068} \frac{0.435}{0.5} = 3.198
\end{eqnarray*}
At the beginning ([$t \leq T_e$]) from Fig. \ref{freq_evolution} we find $\omega_0 = 3$ which is very close to the estimated value ($\omega_0 = 3.198$) from the simple approximate theory. With time the distance between point and patch vortices decreases, thus increasing the natural frequency of the system. The merging is a completely nonlinear process and hence the increase of the natural frequency is also nonlinear.\\

From Fig \ref{freq_evolution} we note that, for $M = 0.5$ the fundamental frequency $\omega_0 = 8$ for the case (b) i.e. \textquotedblleft upto merging\textquotedblright. The timescale associated with this frequency should be $\tau_{0.5} = \frac{2 \pi}{\omega_0} = \frac{2 \pi}{8} = 0.7854$. From previous results, we derive that for $M = 0.05$ the fundamental frequency \textquotedblleft upto merging\textquotedblright~ should be $\omega_0 = 8 \times \frac{0.5}{0.05} = 80$. The corresponding timescale is $\tau_{0.05} = \frac{2 \pi}{\omega_0} = \frac{2 \pi}{80} = 0.0785$. Hence the difference in the timescales will be $\tau_{0.5} - \tau_{0.05} = 0.7854 - 0.0785 = 0.7069$. From Table \ref{merging_time} we see that $T_m(0.5) - T_m(0.05) = 20 - 19.3 = 0.7$ which is very close to $0.7069$. This simple relation is found to hold quite good for all the $M$ values below $M = 0.5$. For even higher $M$ values the contribution from the harmonics and the beats becomes larger and hence the estimate starts deviating from the observed value.\\

With the increase of $M$ the peaks of all the prominent frequencies get broadened and numerous frequencies with relatively much small amplitudes get generated. The power of the peaks of frequencies of kinetic energy
also increases quadratically with $M$ [Fig. \ref{freq}]. This essentially generates several group velocities in the system and thus continuously hammers the vortex crystals with more and more frequently as the compressibility is increased. Many of these frequencies with small amplitudes fall below the cut-off frequency and hence hastens the melting process of the vortex crystals as the compressibility is increased. This can be qualitatively verified from $T_d$ values provided at Table \ref{merging_time} for different $M$.


\section{Analysis of the result for $2 \leq N_{pv} \leq 8$}
\begin{figure}[h!]
\begin{center}
\includegraphics[scale=0.65]{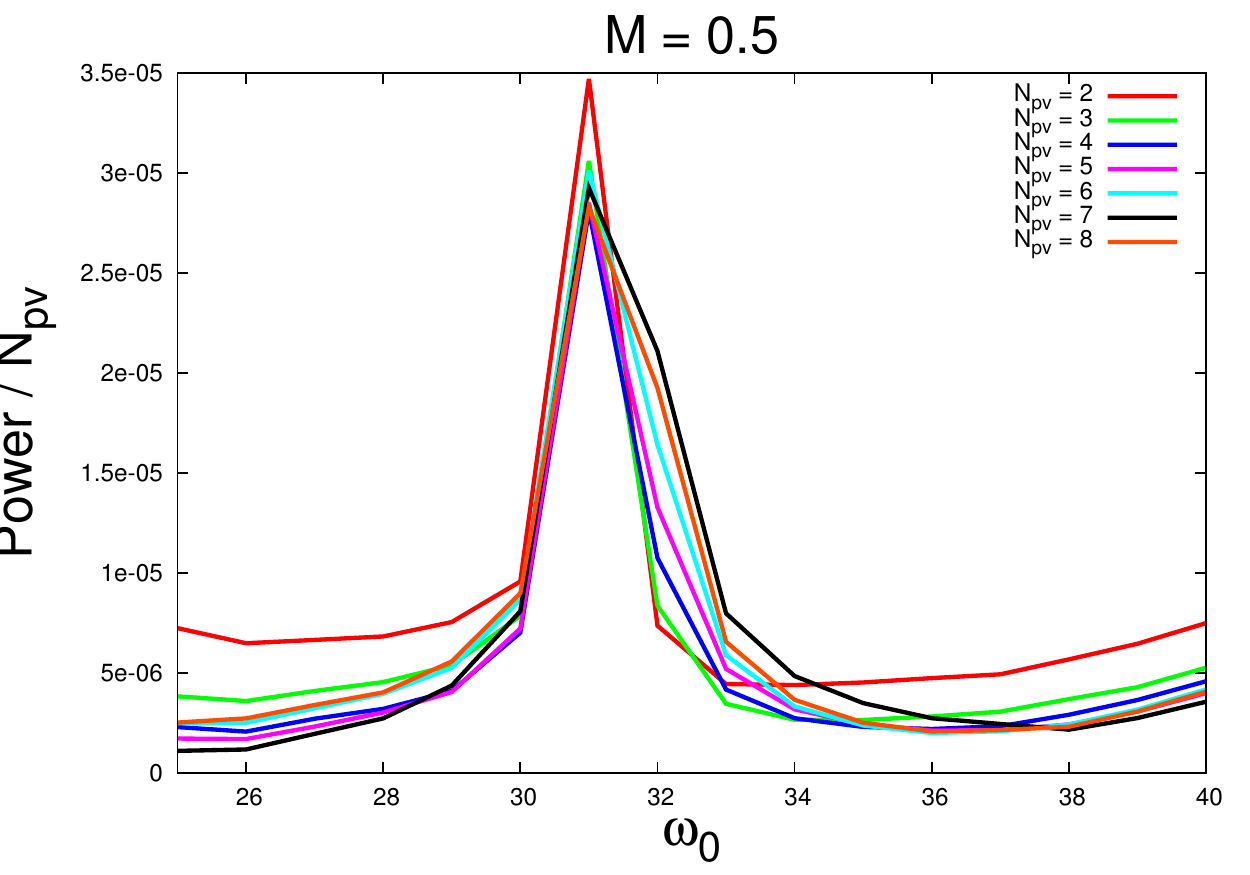}
\caption{(Color online) The fundamental frequency is found to not to vary with changing $N_{pv}$ for $M = 0.5$. The power is found to vary linearly as $N_{pv}$ increases. The other parameters are identical with Table \ref{parameter}}
\label{freq_with_N}
\end{center}
\end{figure}

\begin{figure*}[t]
\begin{center}
\includegraphics[scale=0.33]{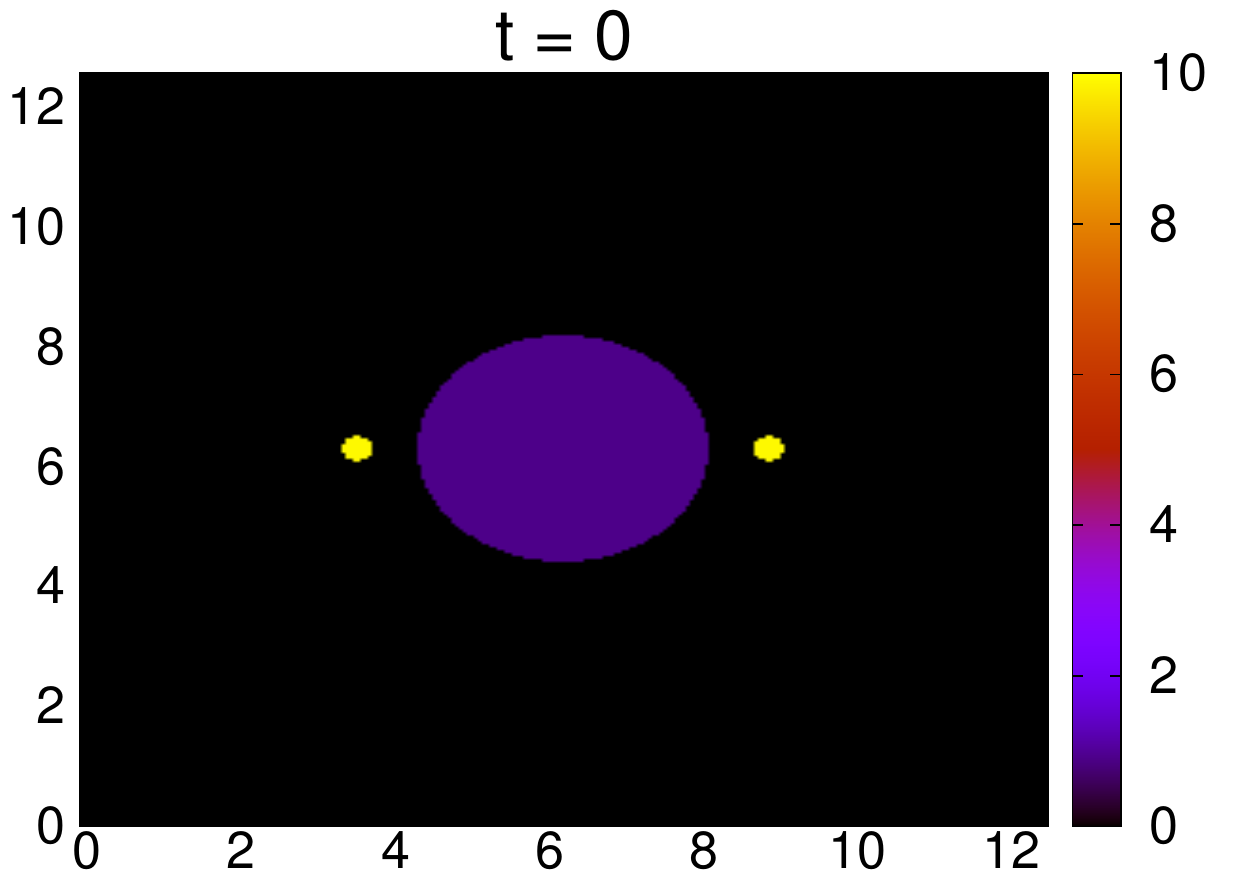}
\includegraphics[scale=0.33]{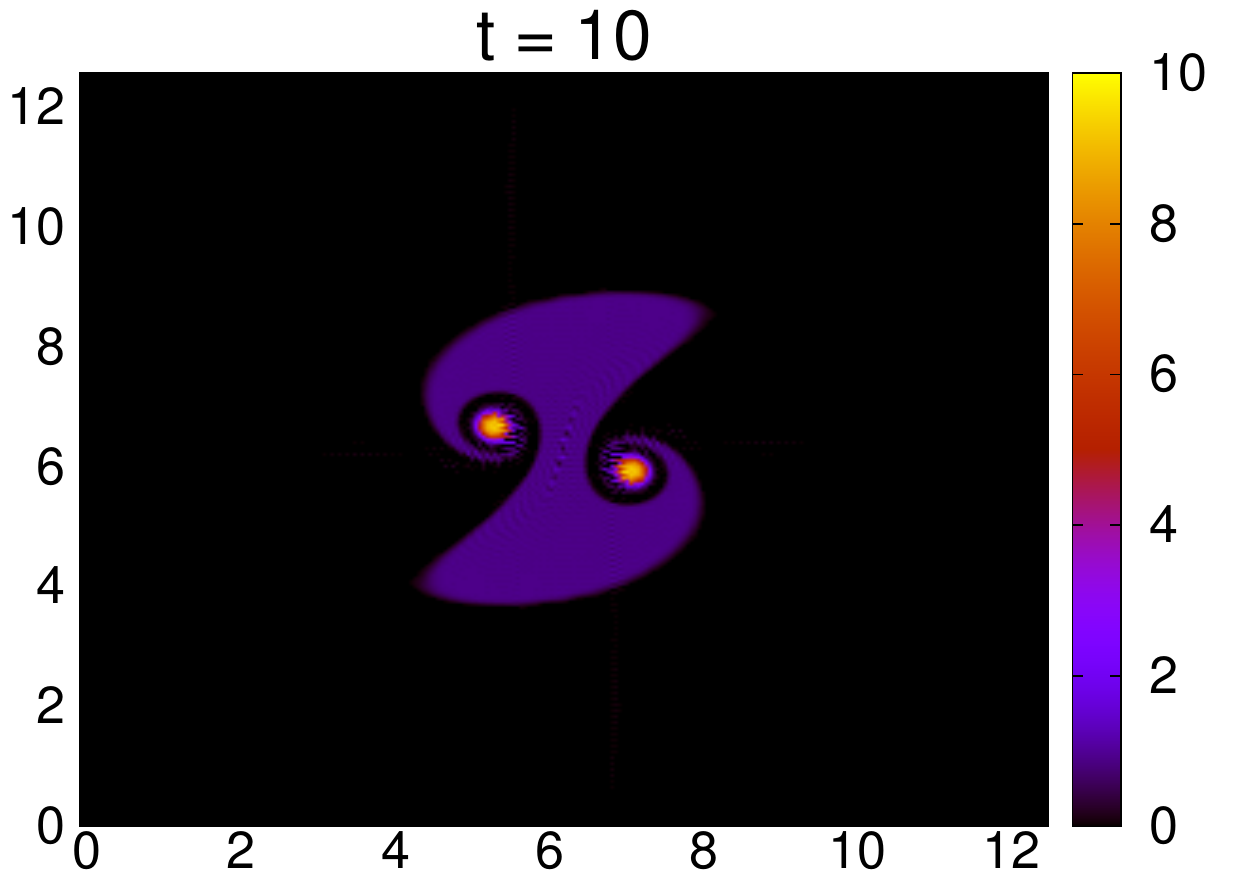}
\includegraphics[scale=0.33]{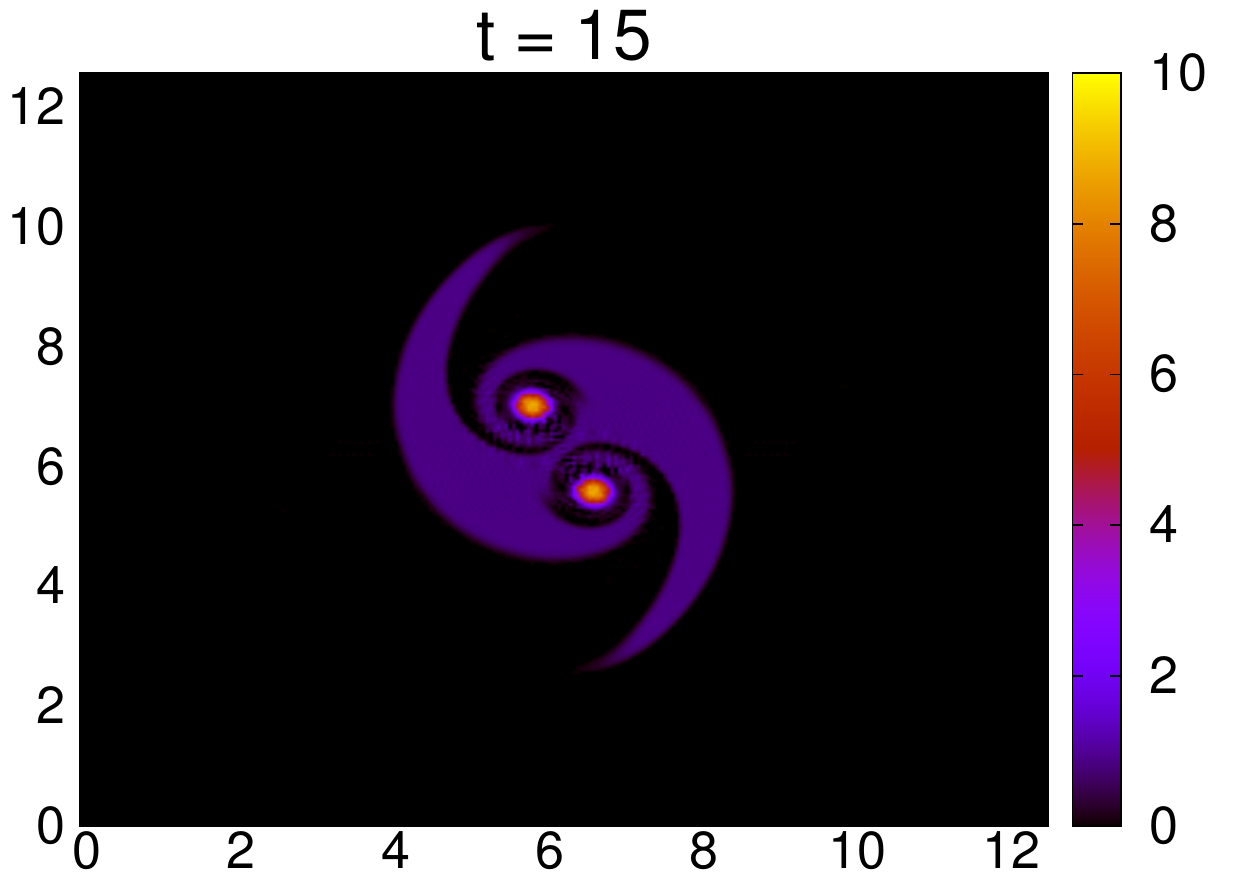}
\includegraphics[scale=0.33]{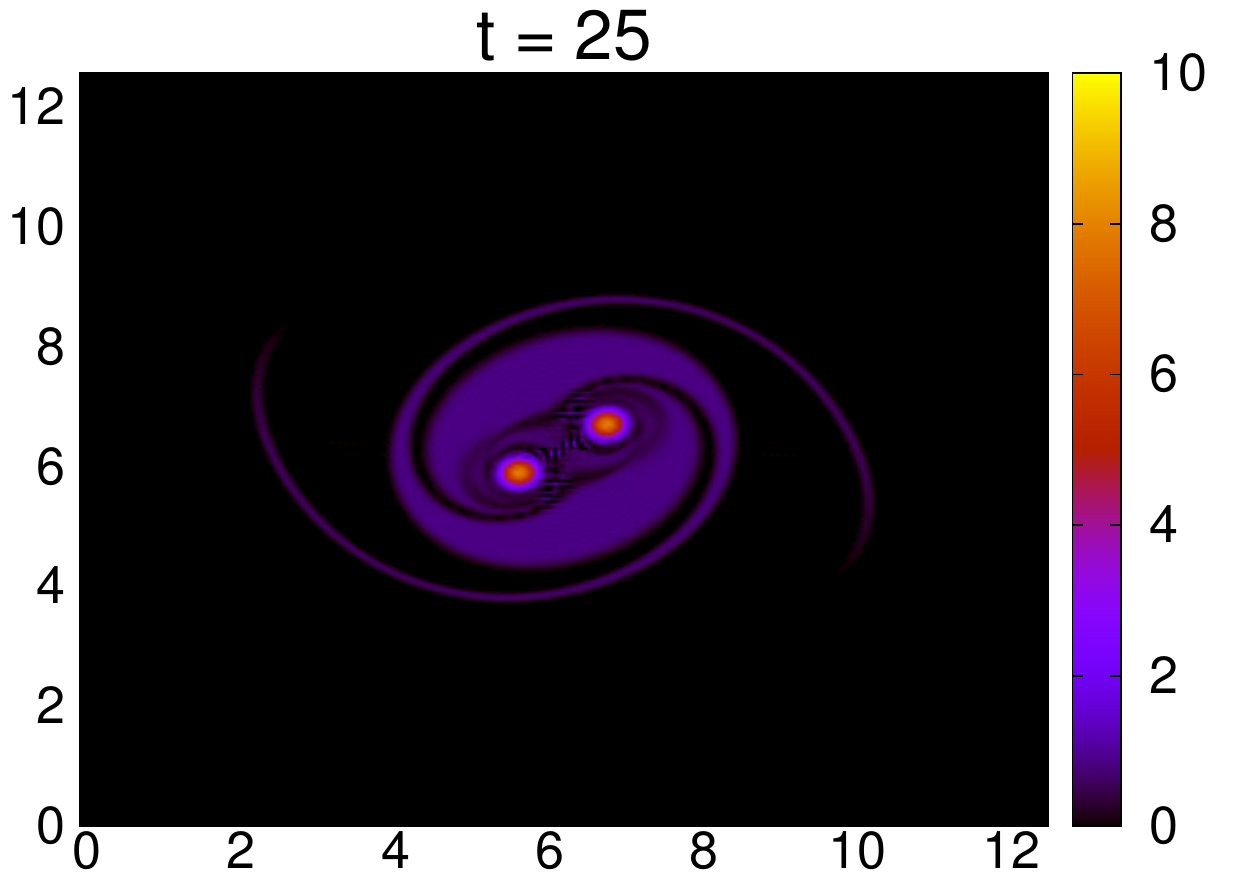}\\

\includegraphics[scale=0.33]{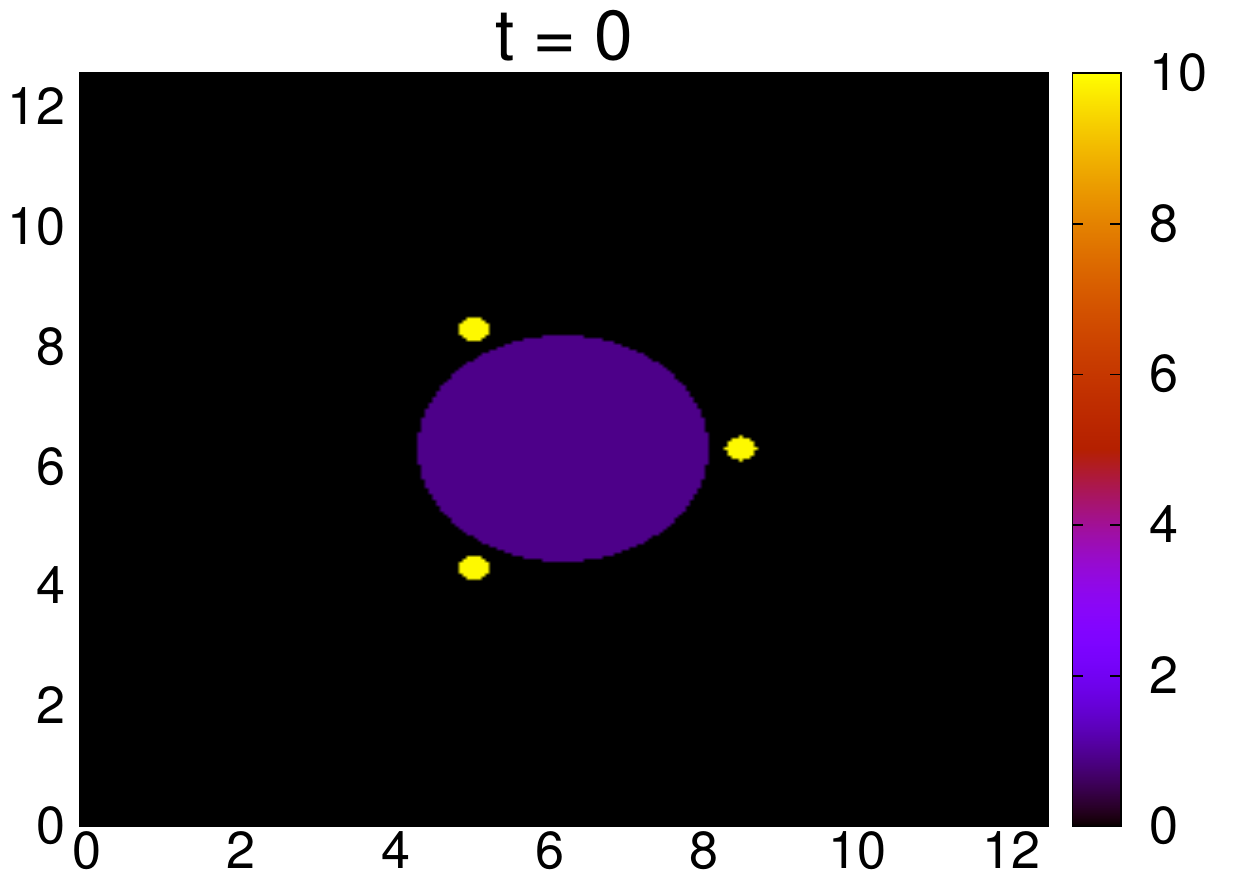}
\includegraphics[scale=0.33]{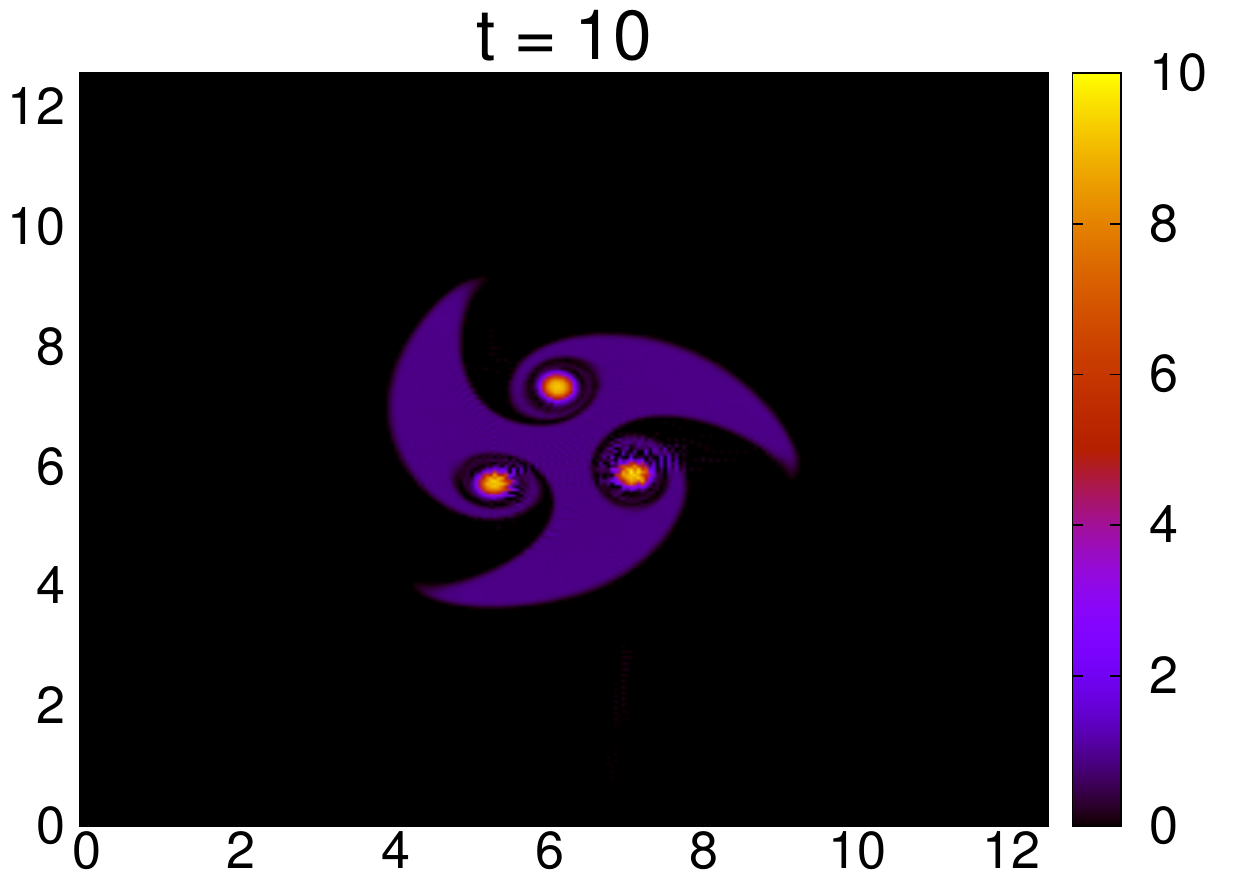}
\includegraphics[scale=0.33]{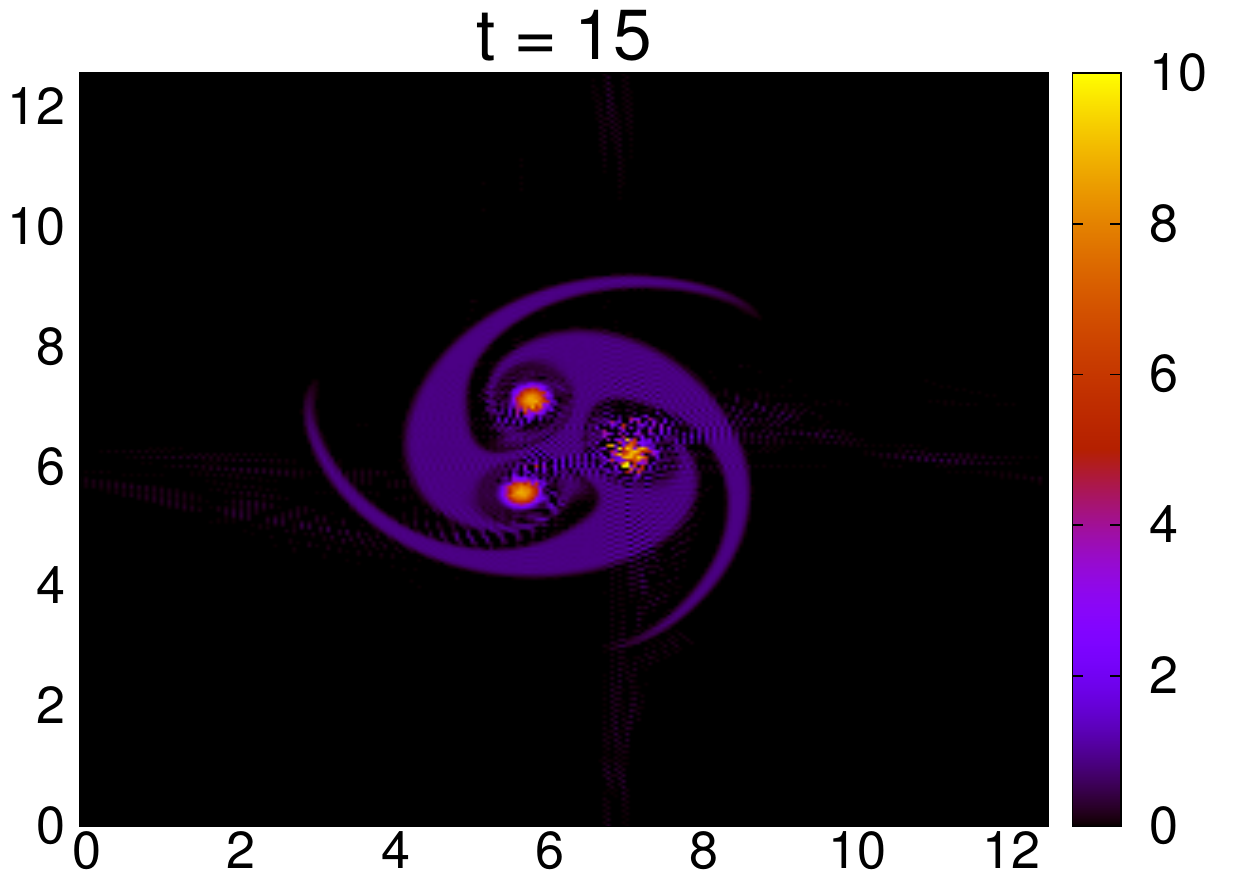}
\includegraphics[scale=0.33]{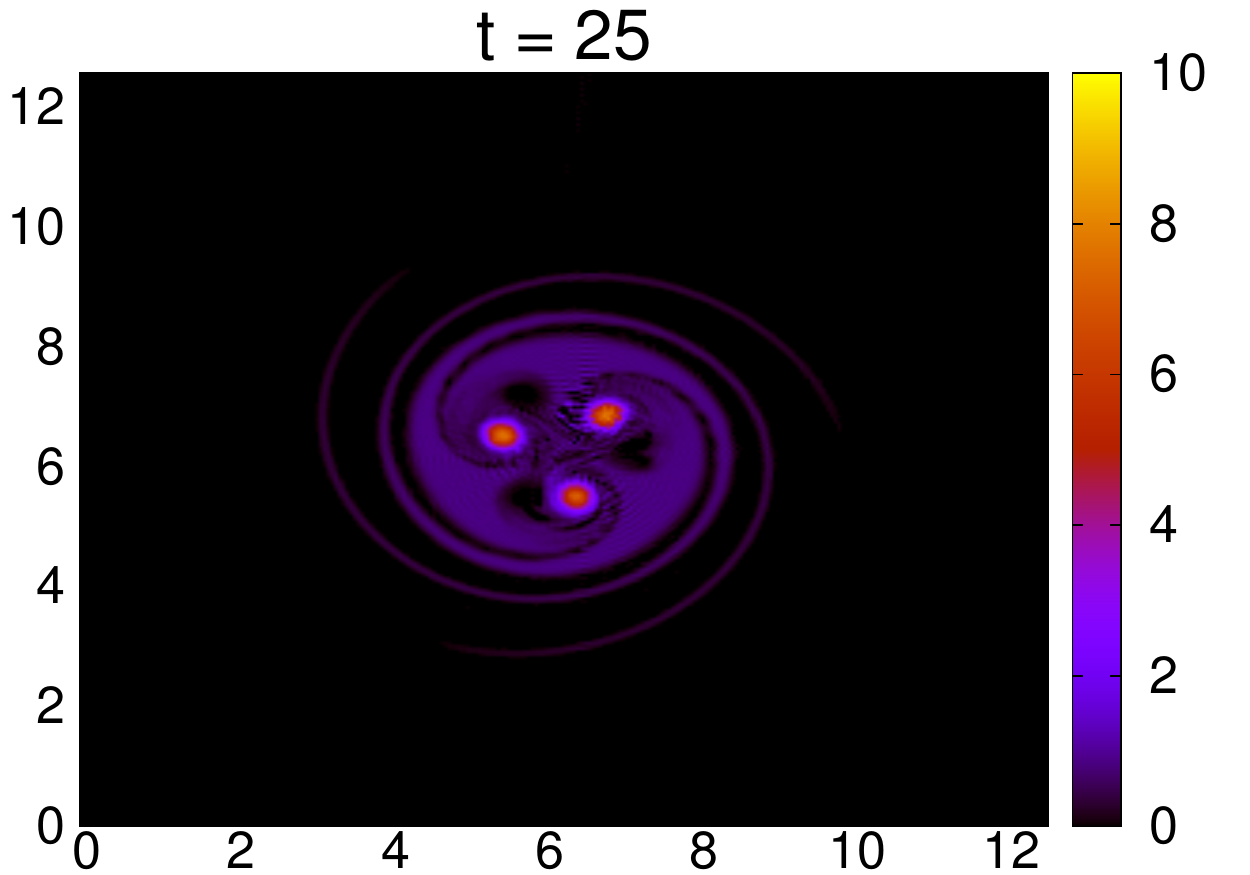}\\

\includegraphics[scale=0.33]{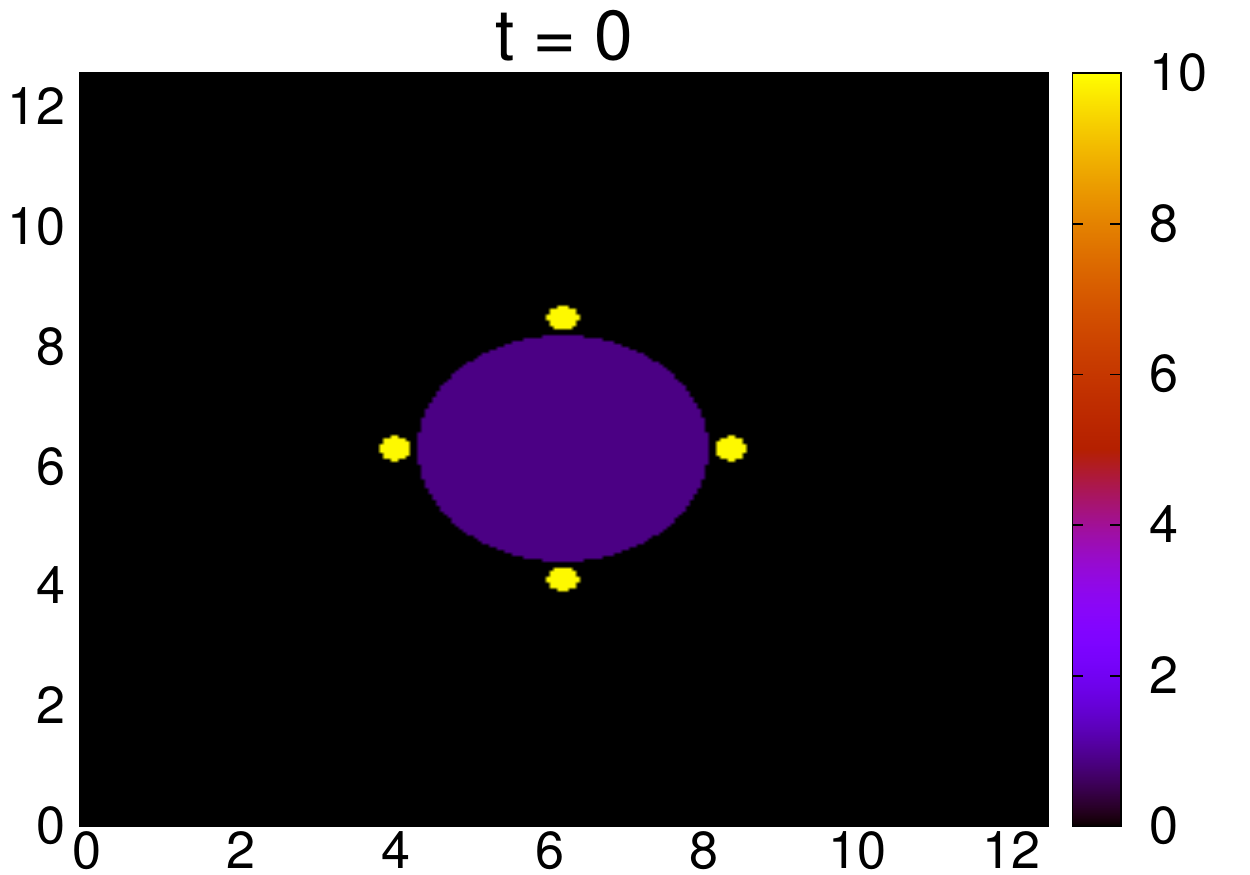}
\includegraphics[scale=0.33]{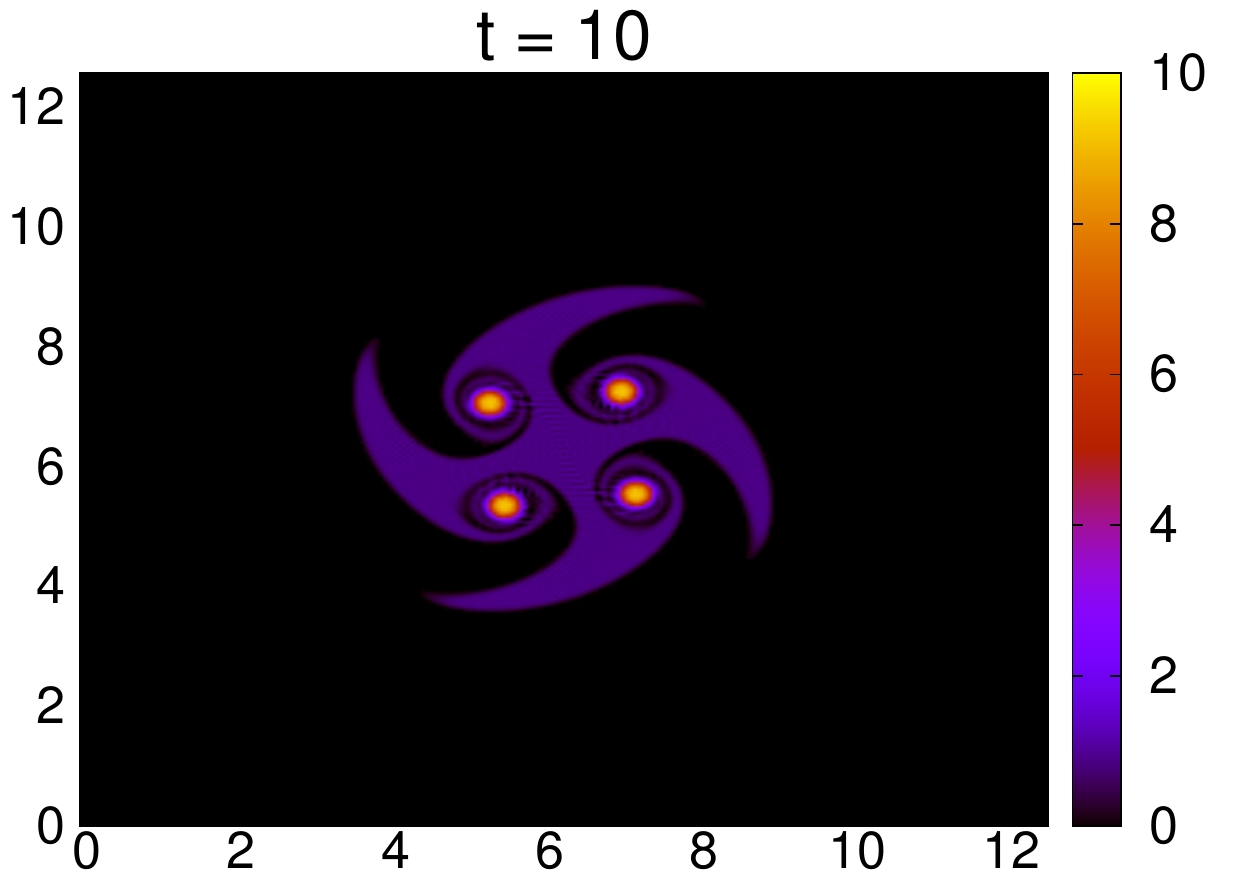}
\includegraphics[scale=0.33]{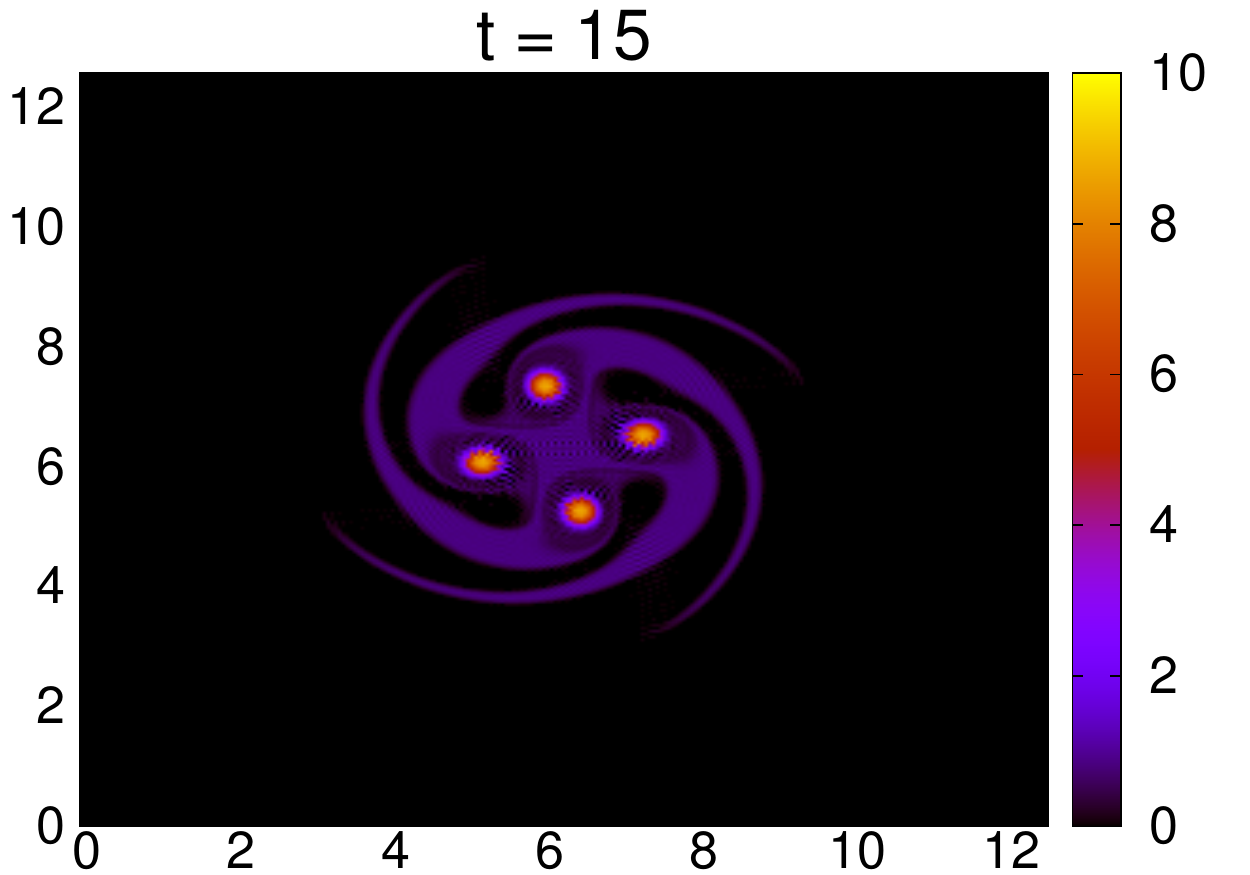}
\includegraphics[scale=0.33]{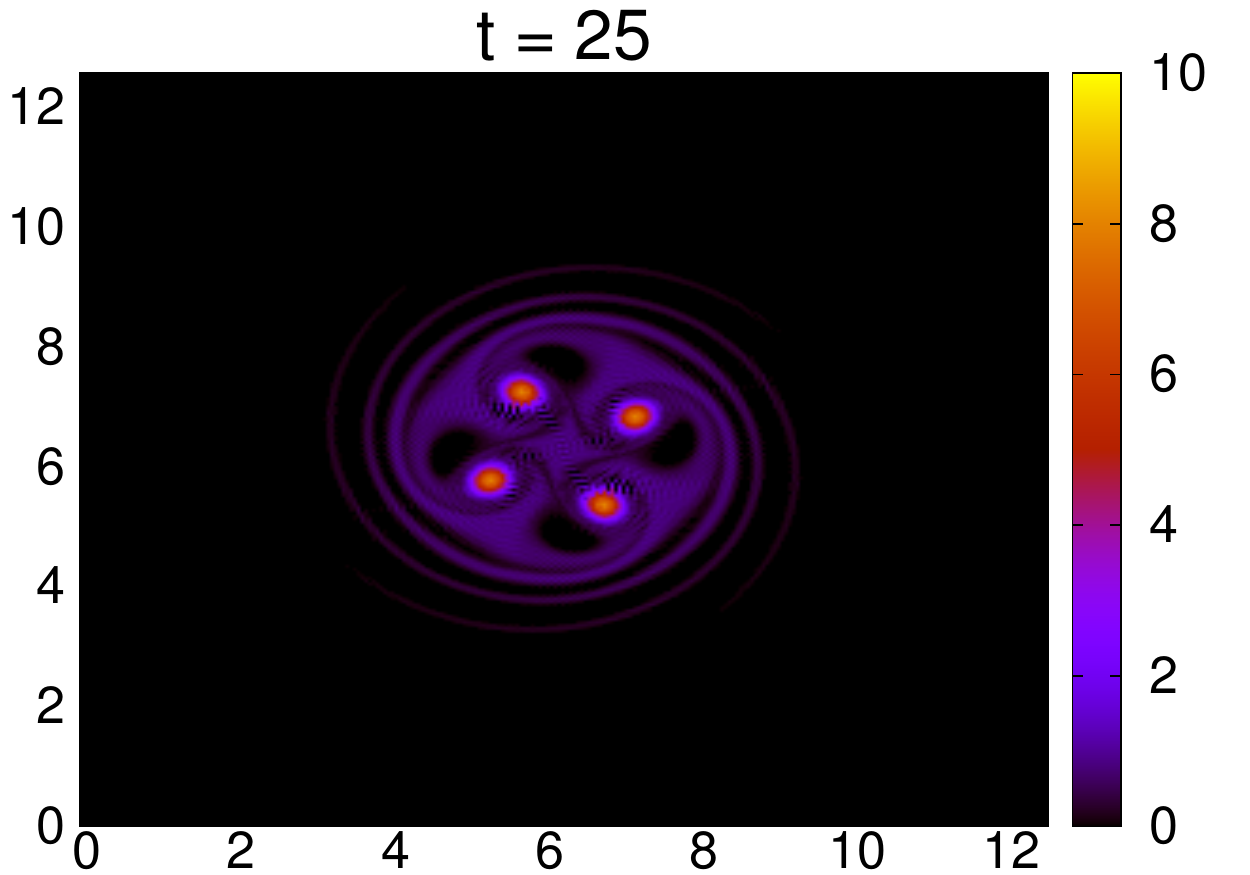}\\

\includegraphics[scale=0.33]{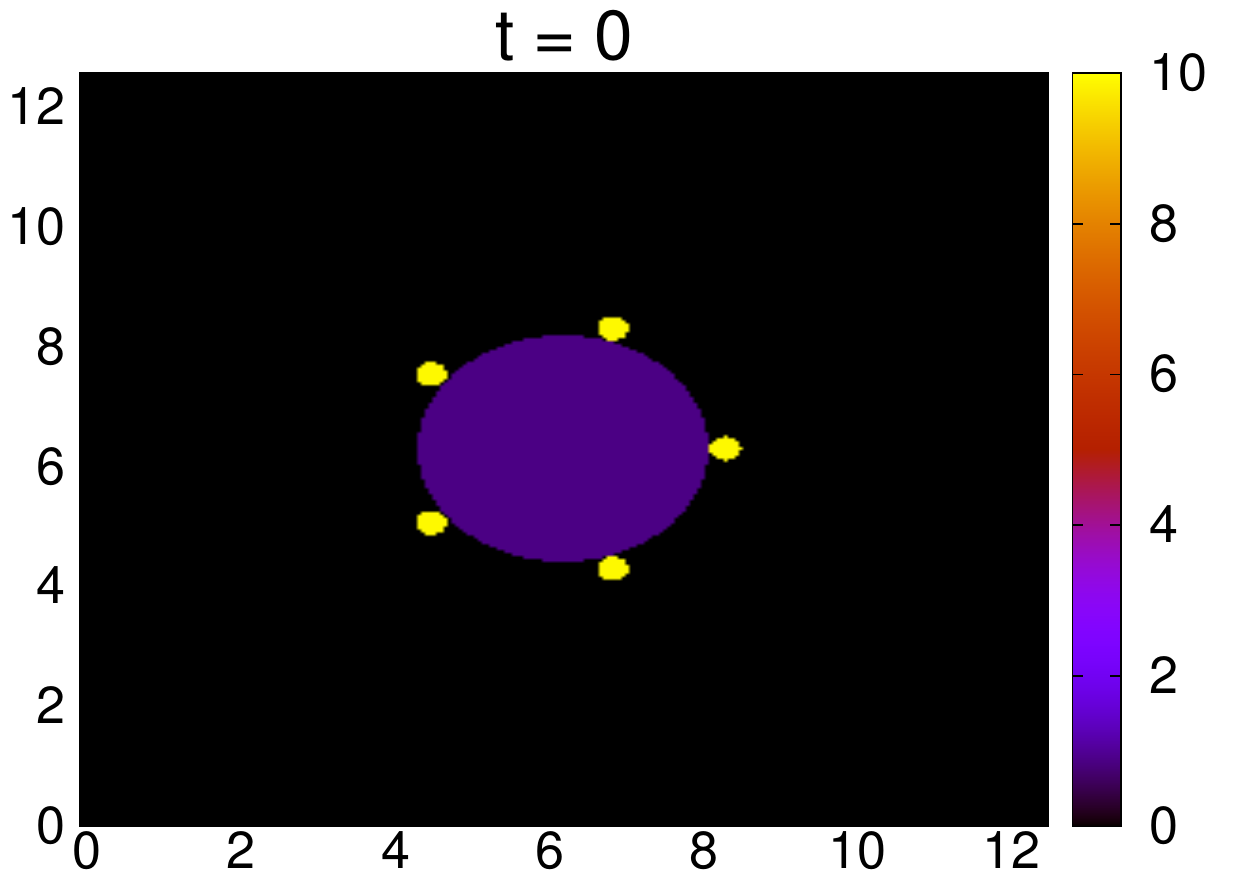}
\includegraphics[scale=0.33]{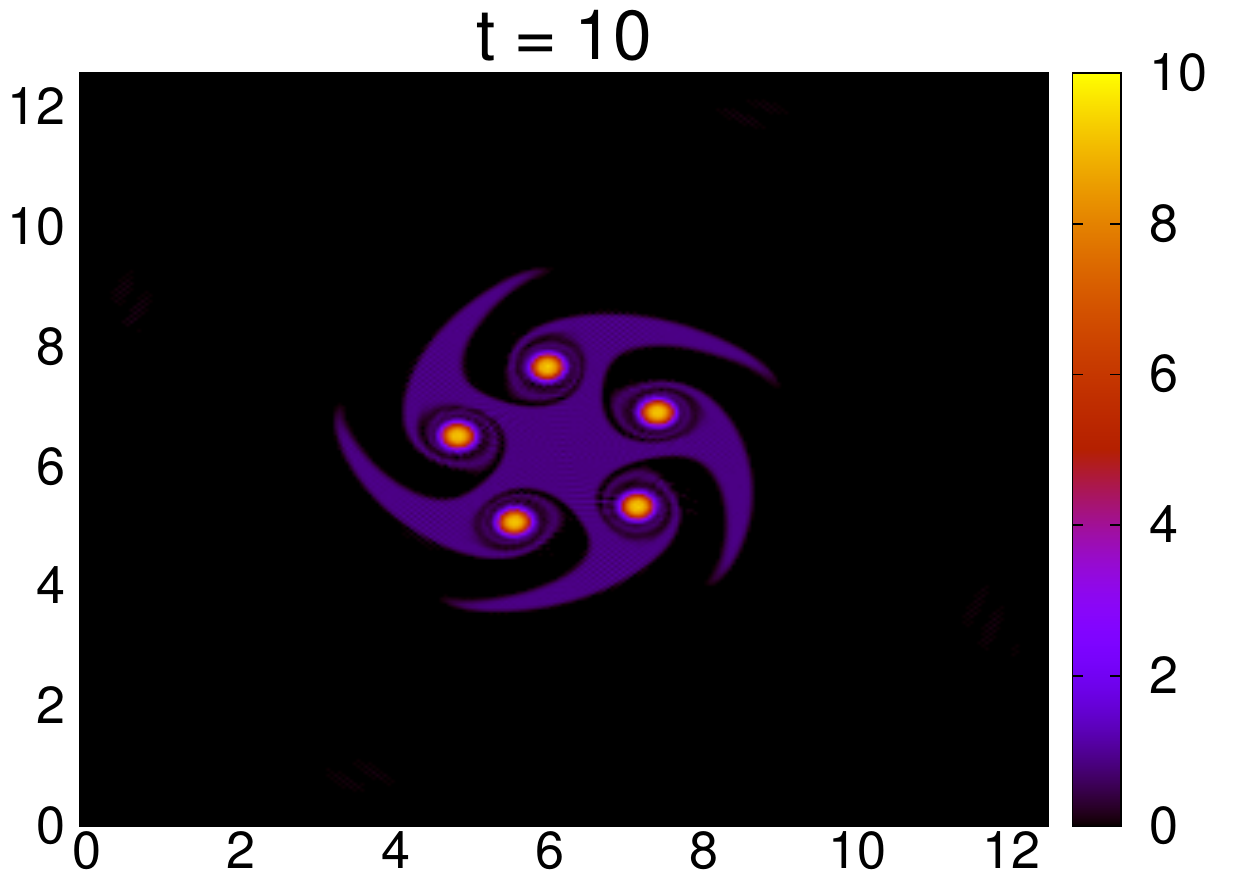}
\includegraphics[scale=0.33]{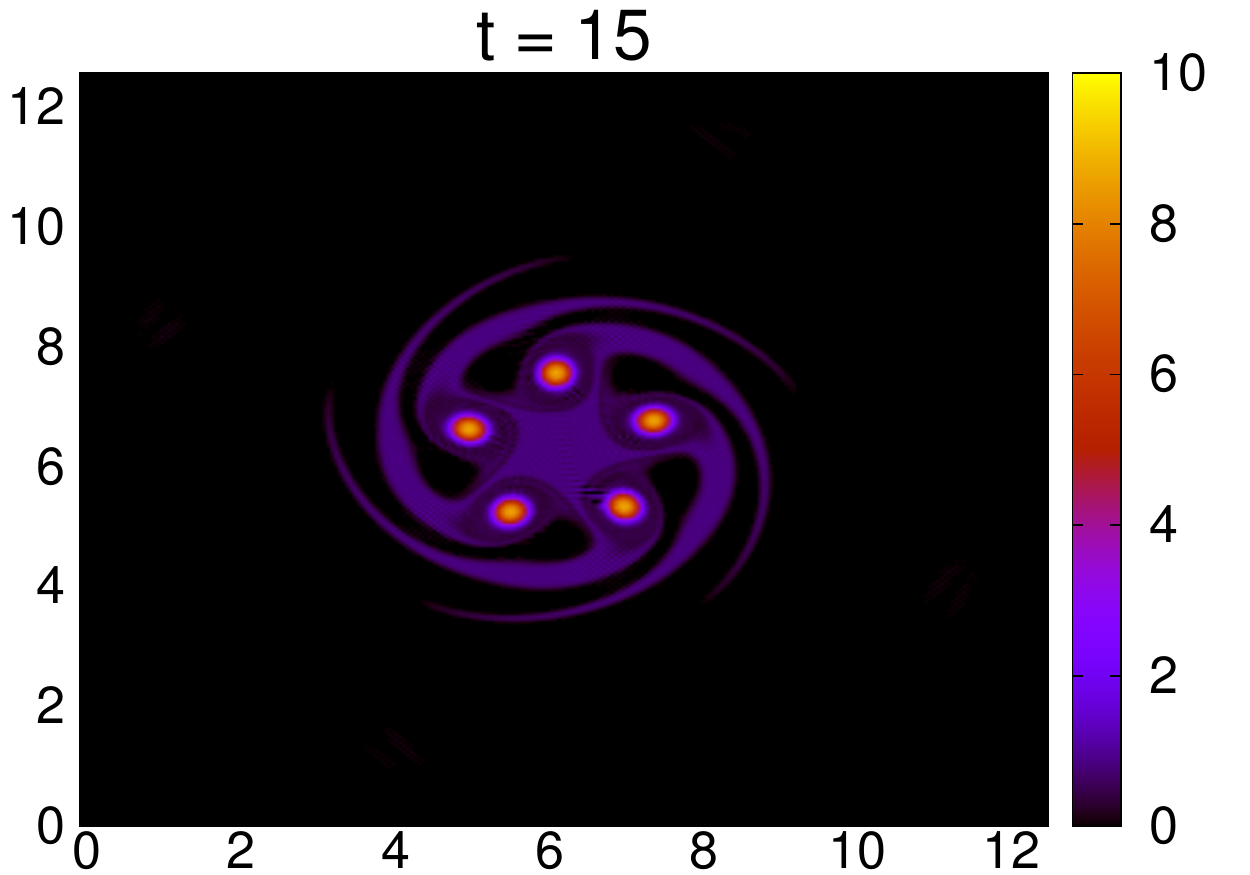}
\includegraphics[scale=0.33]{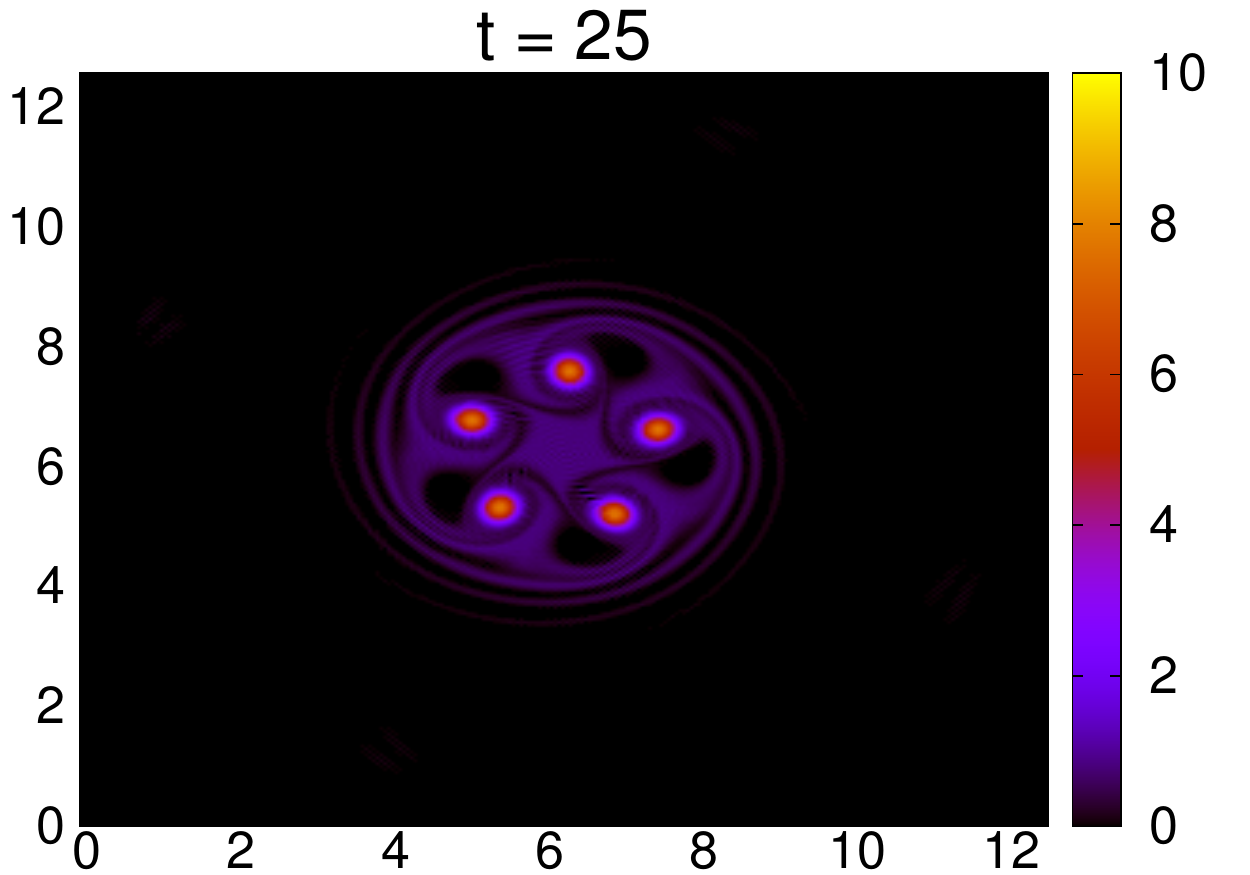}\\

\includegraphics[scale=0.33]{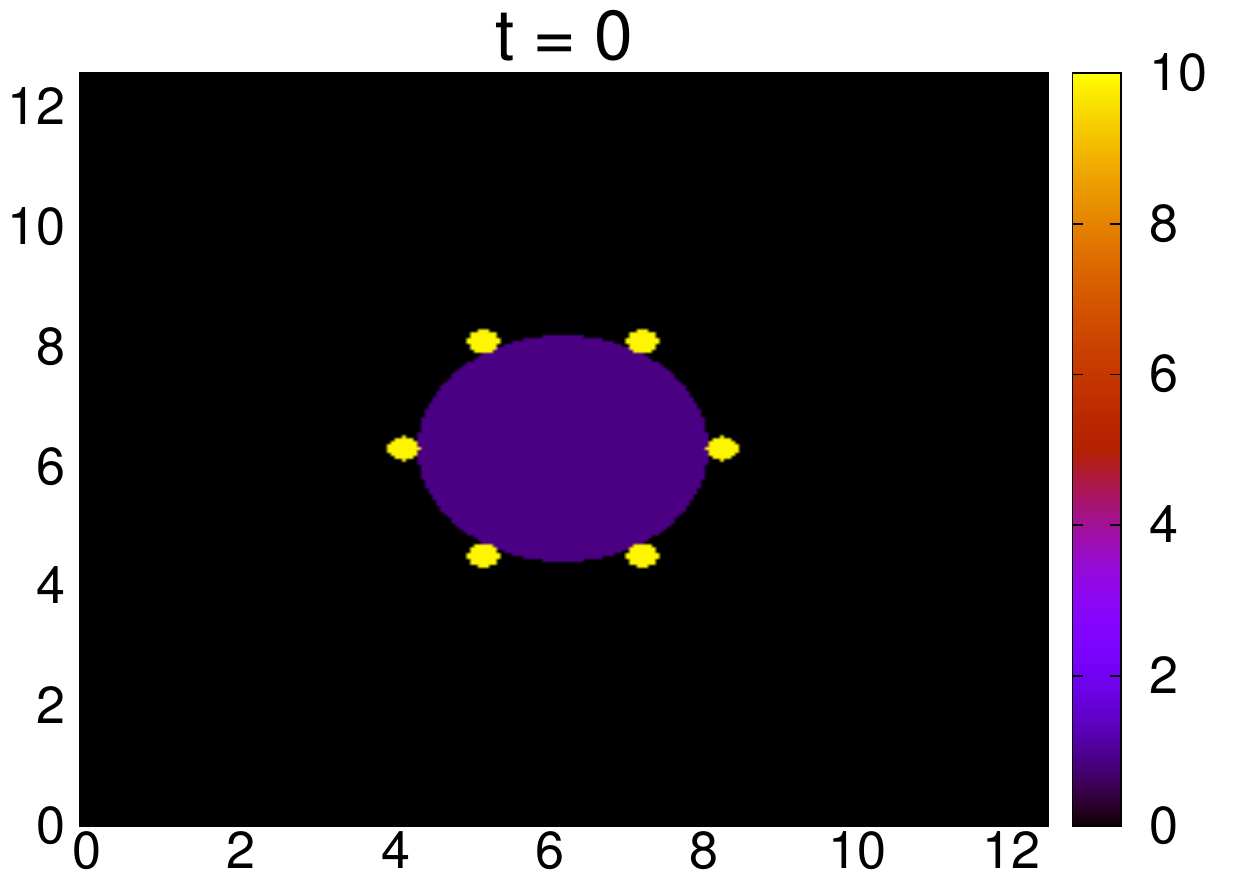}
\includegraphics[scale=0.33]{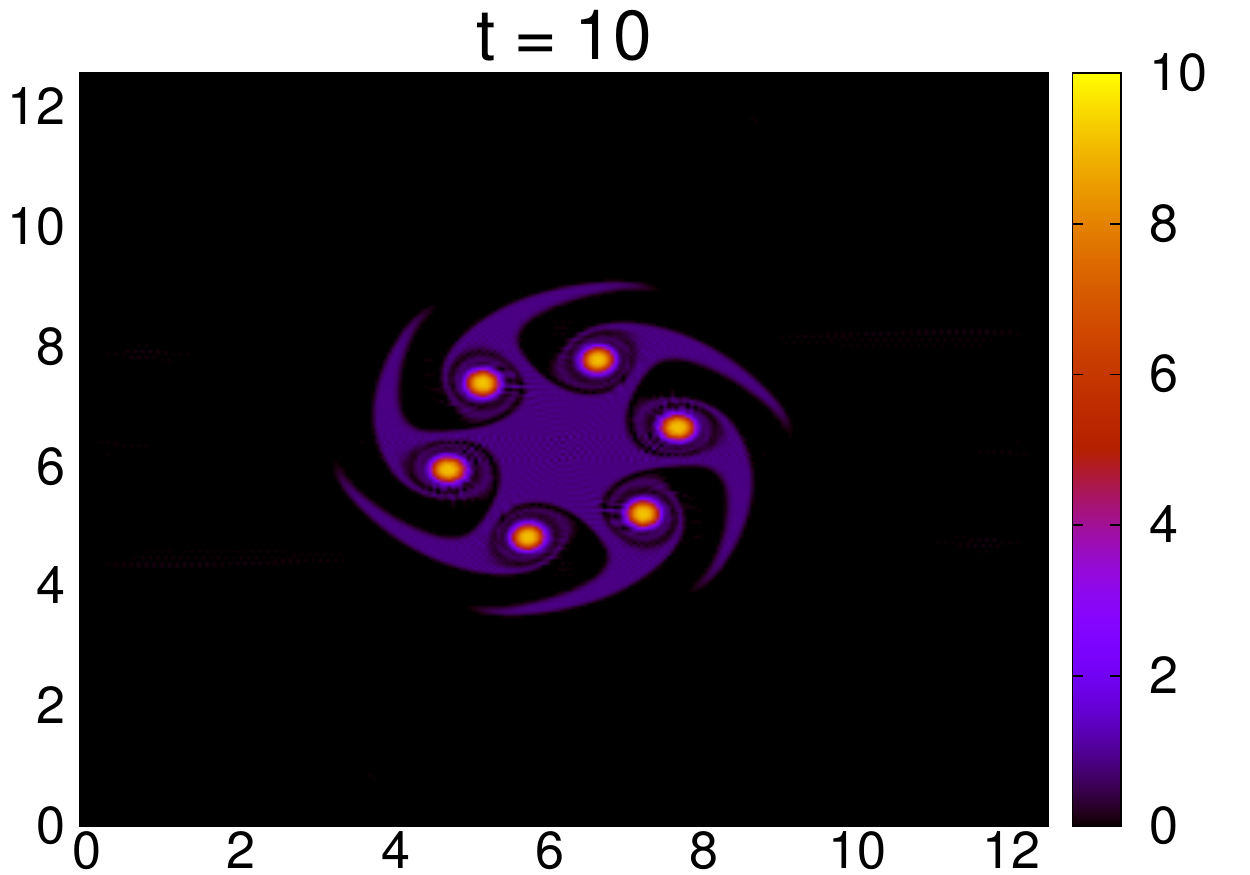}
\includegraphics[scale=0.33]{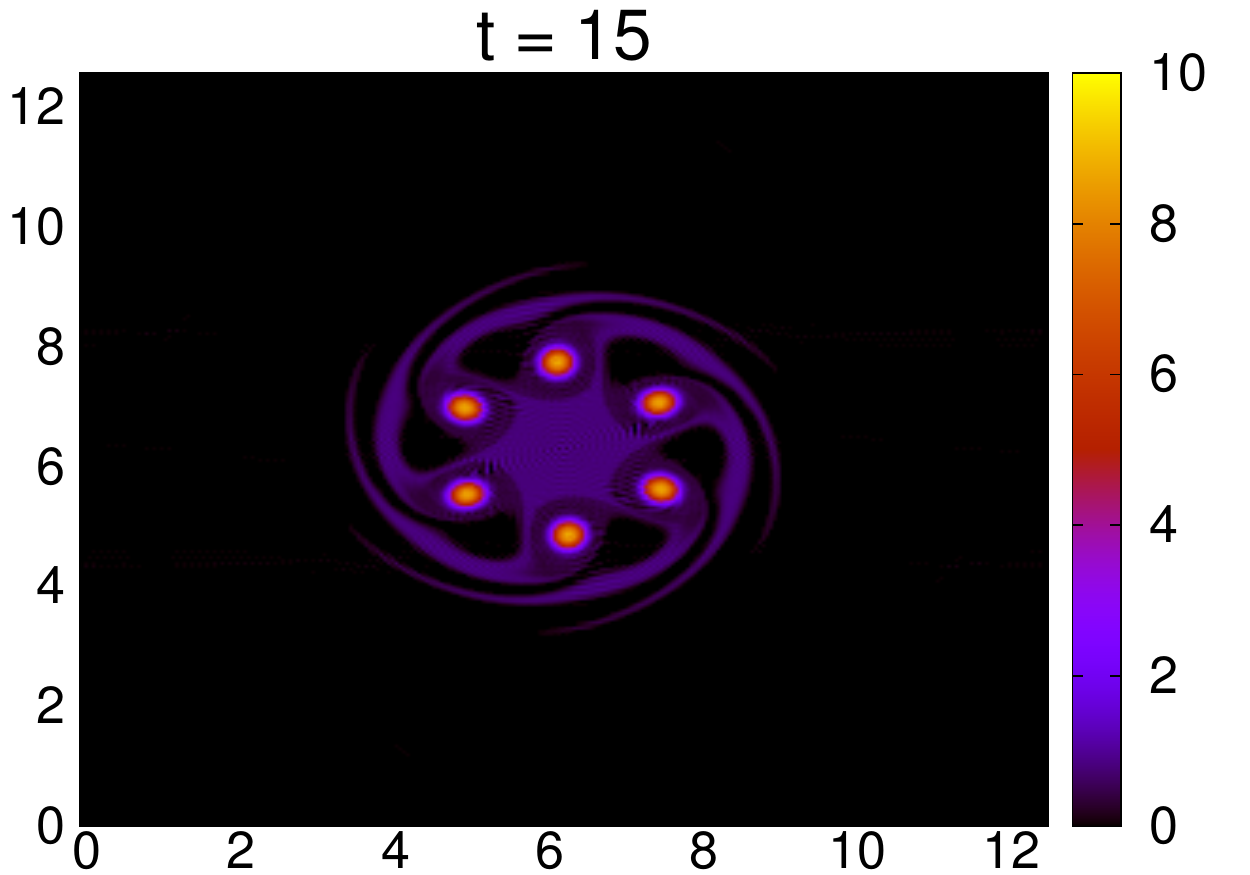}
\includegraphics[scale=0.33]{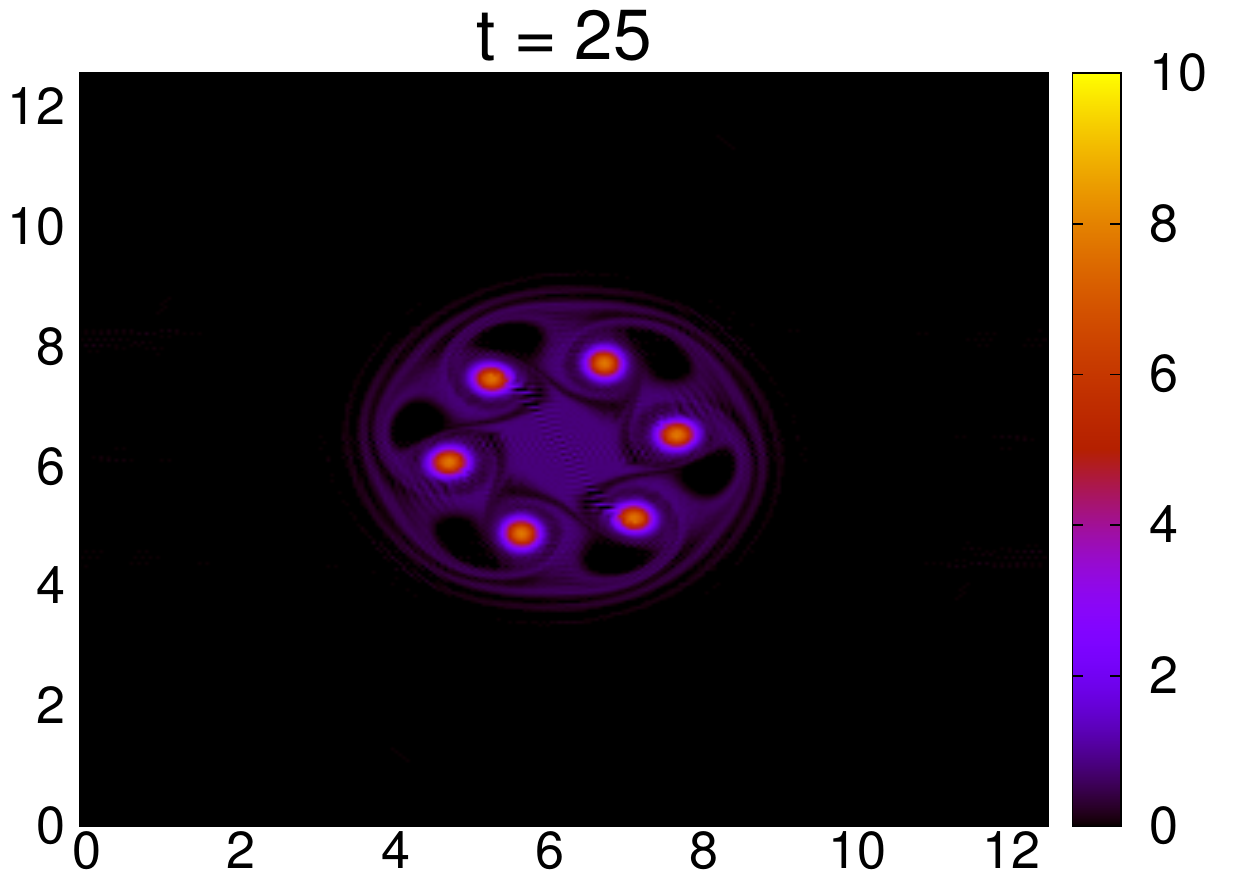}\\

\includegraphics[scale=0.33]{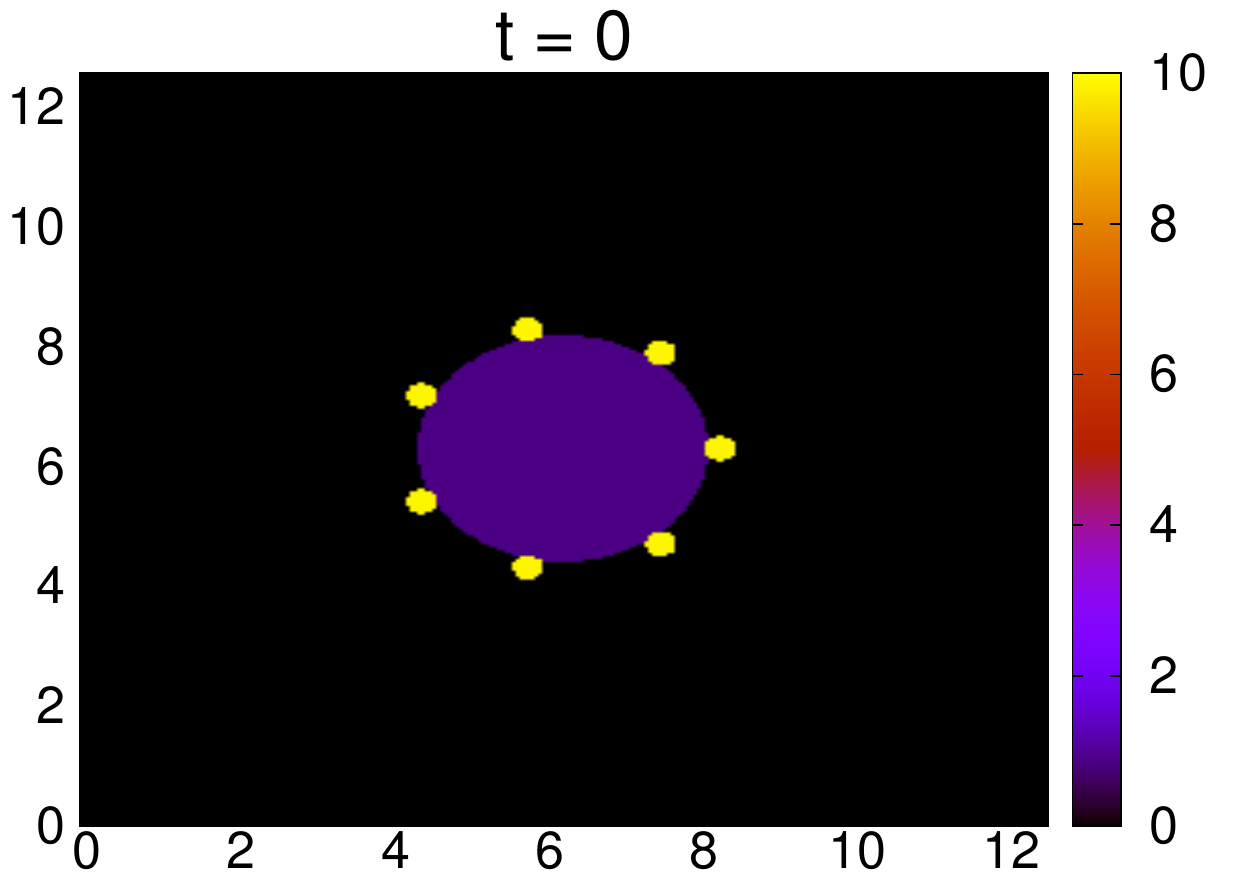}
\includegraphics[scale=0.33]{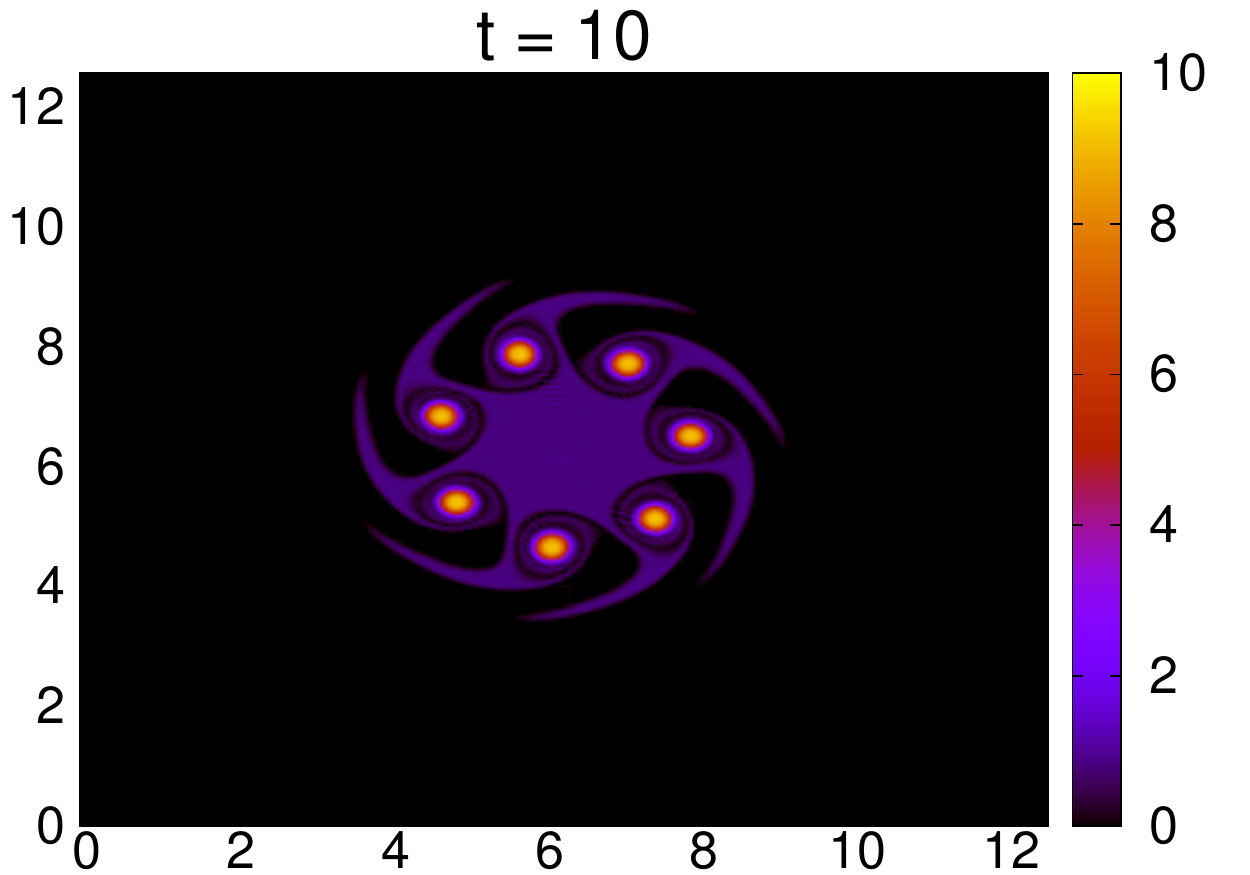}
\includegraphics[scale=0.33]{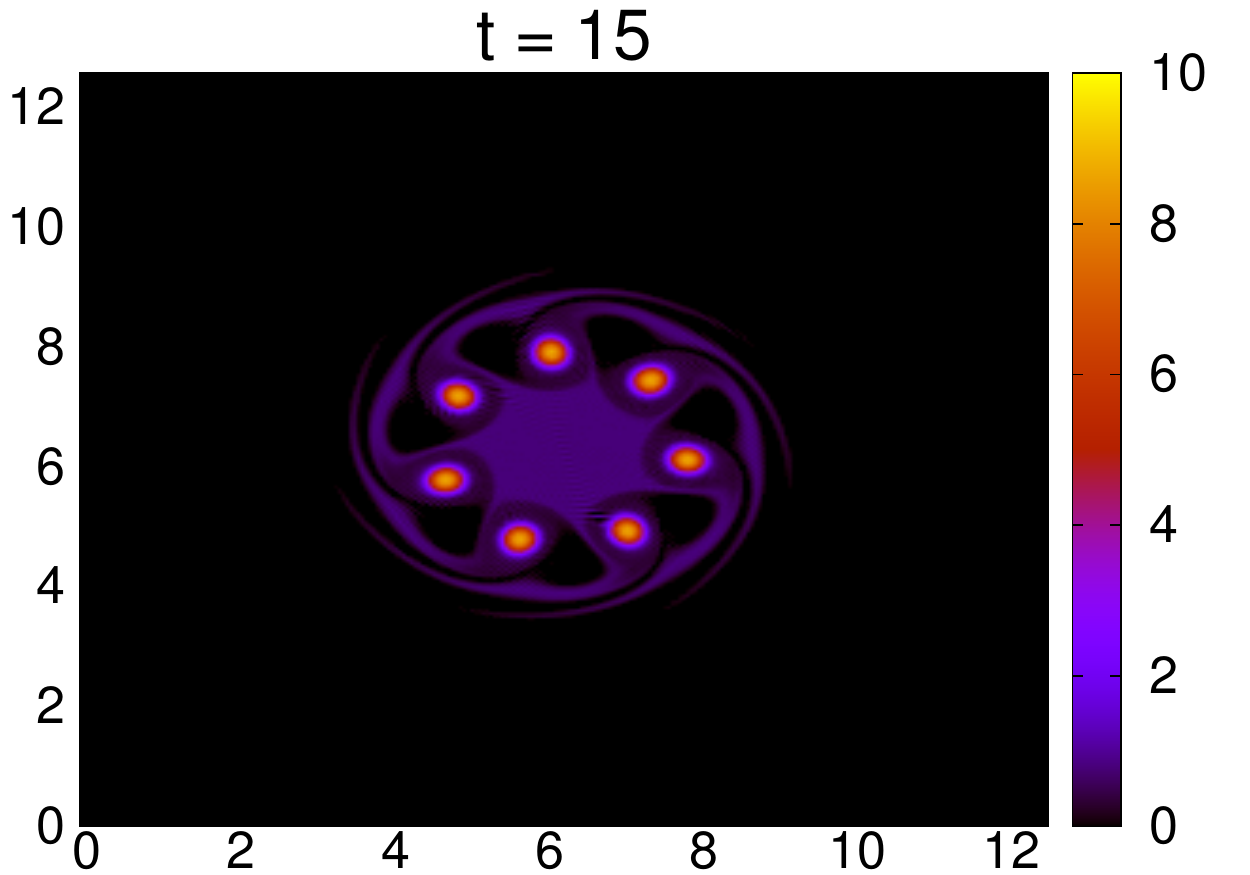}
\includegraphics[scale=0.33]{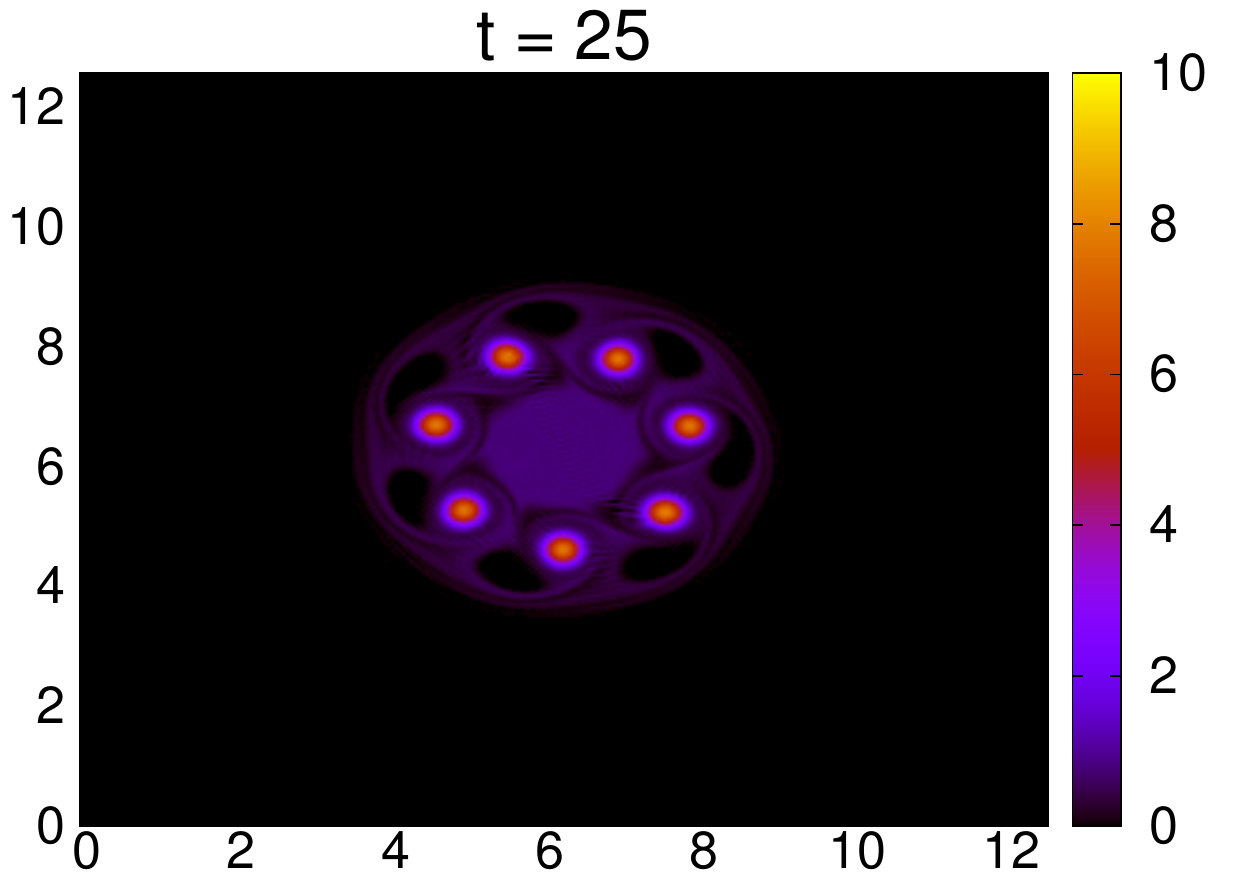}\\

\includegraphics[scale=0.33]{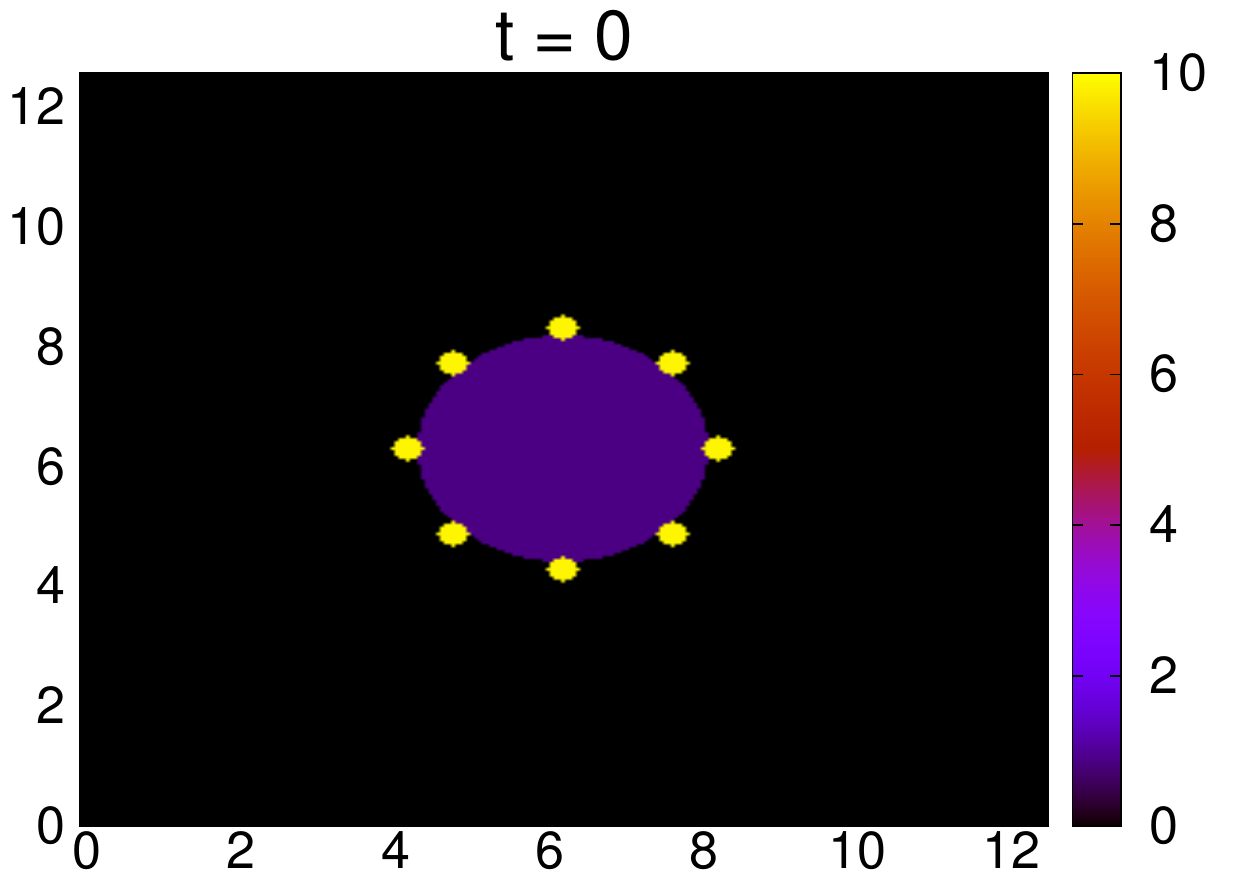}
\includegraphics[scale=0.33]{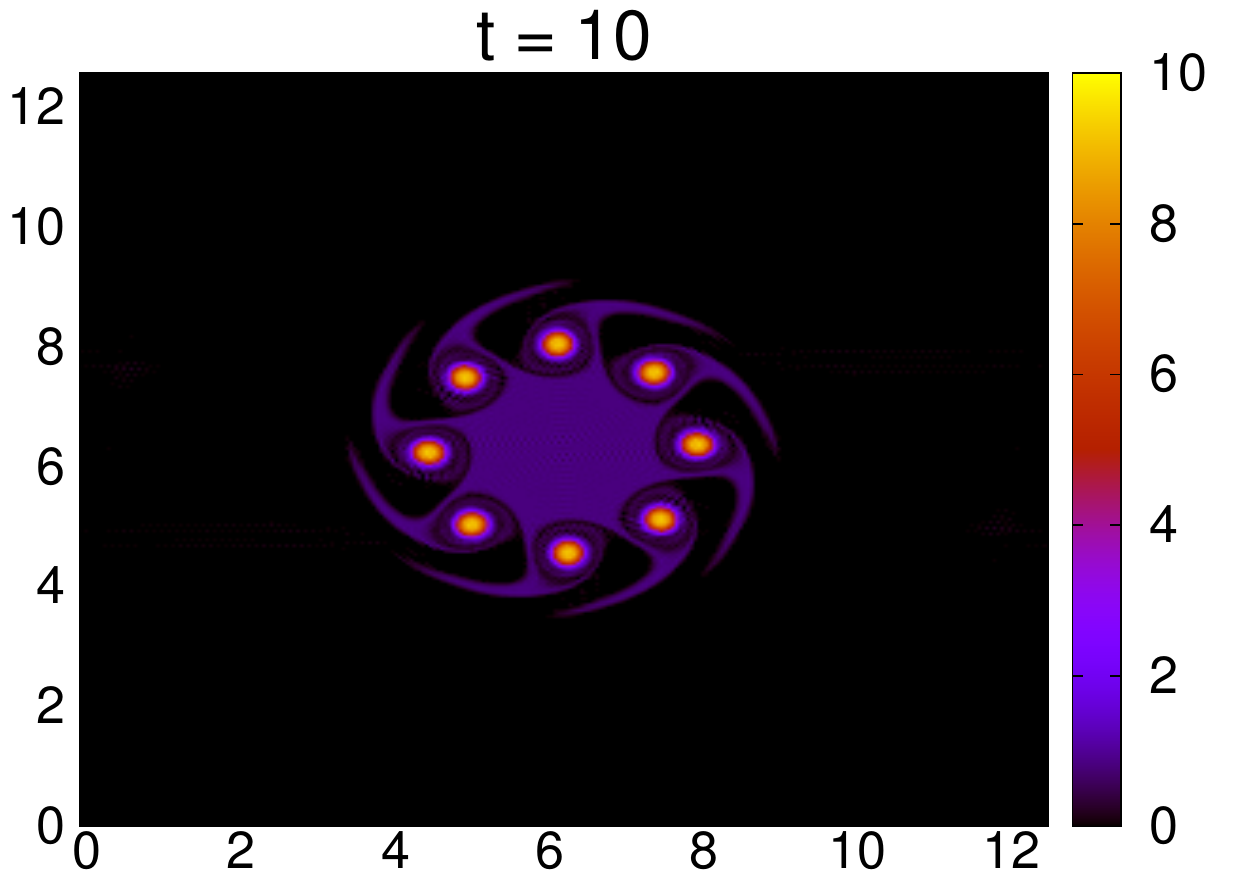}
\includegraphics[scale=0.33]{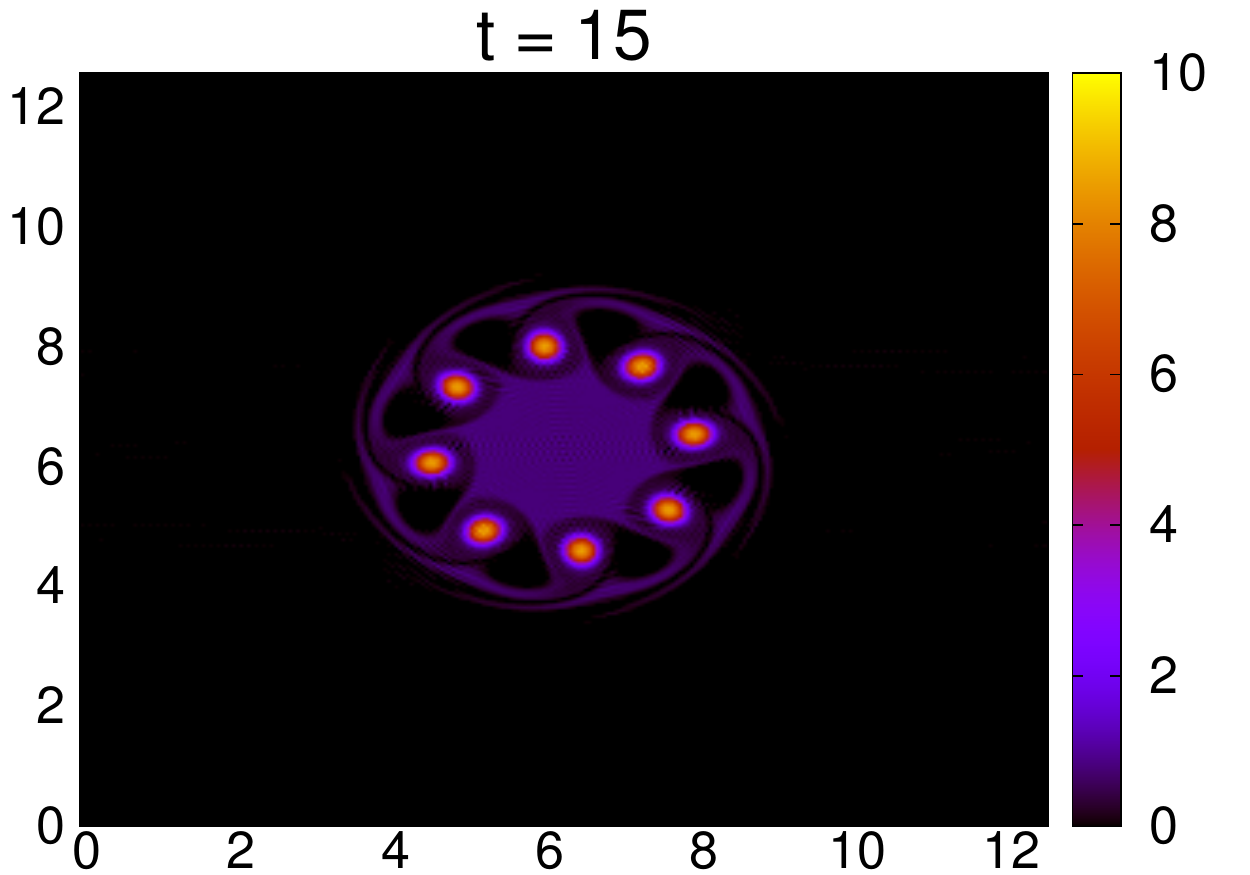}
\includegraphics[scale=0.33]{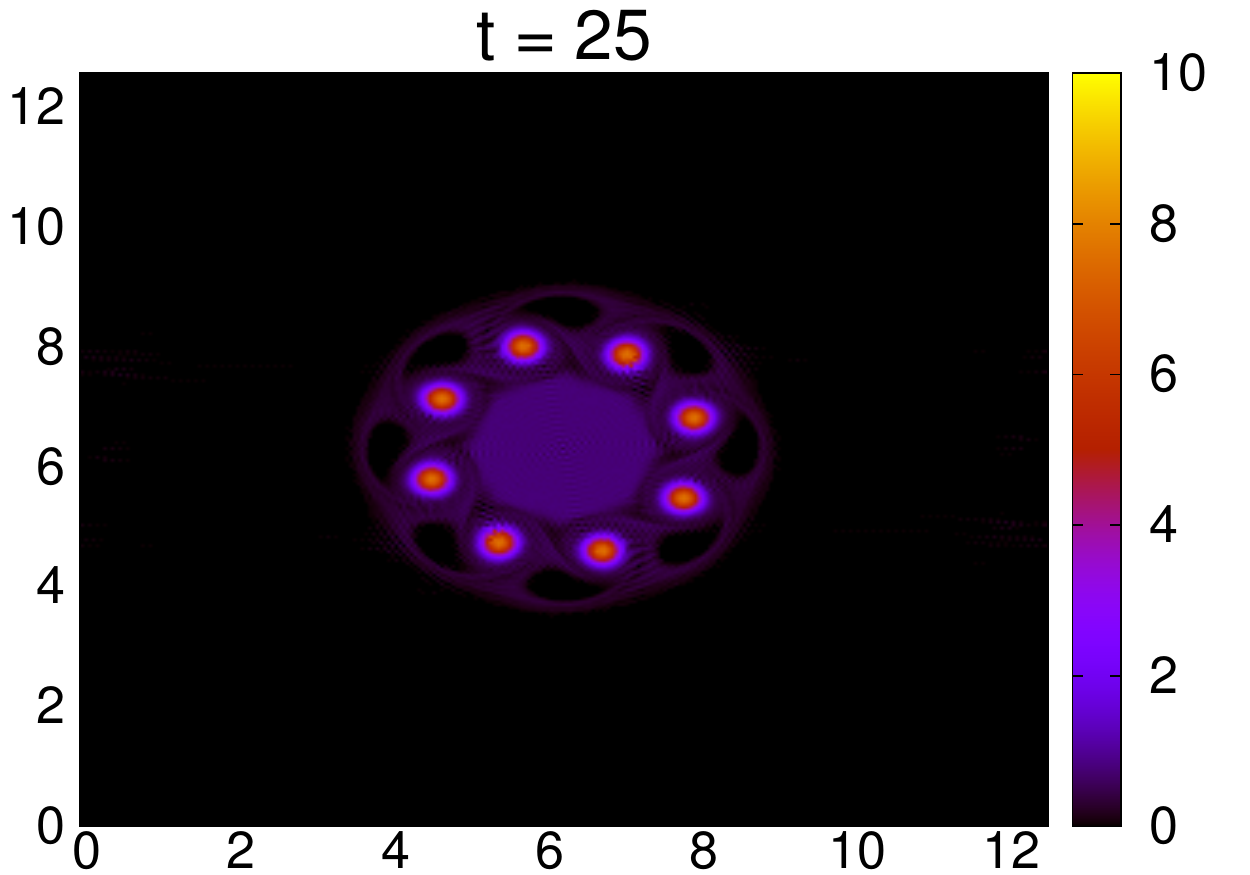}\\
\caption{(Color online) Time evolution of the prearranged vortex merger with $N_{pv} = \{2,3,4,5,6,7,8\}$ and $M = 0.5$ with grid size $256^2$.}
\label{snap_N}
\end{center}
\end{figure*}

Now, we perform simulation for $2 \leq N_{pv} \leq 8$. As previously found \cite{ganesh:2002}, the $R_{pv}$ needs to be tuned to induce the instabilities and generate the elongated fingers at proper places. We keep $\frac{R_{pv}}{R_{pa}} = \sqrt{\frac{N_{pv}}{N_{pv}-1}}$ to guarantee merging. We have also found that changing $R_{pv}$ by a small value (at least upto $0.21 R_{pa} \frac{L}{2}$) around the prescribed value does not affect any of the frequencies obtained earlier $\left(\sqrt{\frac{N_{pv}}{N_{pv}-1}}\right)$. This indicates that, there is a band width of radial location for a fixed $N_{pv}$ value for which this merging process can take place. The distance has to guarantee the excitation of the Kelvin waves around the patch vortex even in the presence of viscosity. We run the simulation for a constant Mach number (M) = 0.5. We have found that by changing the number of point vortices $N_{pv}$ the fundamental frequency as well as its other harmonics and the beats does not change but the power changes. The power of kinetic energy is found to vary linearly with $N_{pv}$ (Fig.\ref{freq_with_N}). It indicates that the fundamental frequency arises due to the interaction between the patch and the point vortex only and not due to the interaction between the point - point vortex. The linear scaling of power with $N_{pv}$ also supports the indication. We also provide some snapshots of compressible vortex mergers with $N_{pv} = 2,3,4,5,6,7,8$ for the identical parameter regime mentioned above [Fig \ref{snap_N}].\\
 
\section{Summary and discussion}
In summary we have numerically time-evolved a prearranged set of point-like vortices centered around but outside a circular patch vortex and observed its merging and the formation of holes and quasistationary vortex crystals.\\

We have found that the compressibility effect induces a natural mode ($\omega_0$) to the vortex dynamics along with its harmonics and the beats with the harmonics. The power of kinetic energy is found to scale as $M^{-2}$ where $M$ is the Mach number. For lower $M$, this fundamental frequency ($\omega_0$) is identified to affect the merging time as well as the vortex crystal-hole pattern formation time.\\

Unlike the report earlier \cite{ganesh:2002}, the vortex crystals are found to melt faster as the compressibility is increased. As the Mach number is increased, the generation of multiple frequencies with comparatively less power are primarily believed to affect the sustenance of the vortex crystals.\\

We have verified that the natural frequency does not change with the number of point like vortices. Also we have seen that the power of kinetic energy scales linearly with the number of point vortices. Thereby we have inferred that this natural frequency arises due to the interaction between the patch and the point vortices. \\

Magnitude of the initial conditions (eg. density, temperature, entropy) are known to affect the dynamics of the fluid \cite{terakado:2014, bayly:1992} and hence it would be interesting to investigate the scaling of the quantities found in this paper for different initial conditions of the fluid.\\


In the present work, a detailed comparision with two different spectral codes has been presented. However, for Mach flows, it would help if such a comparative study is performed using shock capturing numerical schemes. This will be addressed in a future communication.\\

Our study extends the direct parallel between two dimensional turbulence and magnetised plasma column under strong axial magnetic field as pointed out earlier \cite{ganesh:2002} to the compressible regime which is more realistic in any laboratory experiment, though the exact isomorphism between two dimensional ideal incompressible fluid dynamics and magnetised two dimensional plasma is broken.


\section{Acknowledgement}
All works have been performed at Uday and Udbhav cluster at Institute for Plasma Research. The authors thank Abhijit Sen, IPR for his valuable comments on this manuscript. RM thanks Samriddhi Sankar Ray at International Center for Theoretical Sciences, India for his initial help regarding pseudo spectral simulation. RM also acknowledges the support from ICTS program: ICTS/Prog-dcs/2016/05.



\bibliography{biblio}

\end{document}